\documentclass[%
reprint,
amsmath,amssymb,
aps,
prl
]{revtex4-1}

\usepackage{tabu}
\usepackage{graphicx}
\usepackage{epstopdf}
\usepackage{dcolumn}
\usepackage{bm}
\usepackage[breaklinks=true]{hyperref}
\hypersetup{colorlinks=true, citecolor=blue, filecolor=blue, linkcolor=blue,urlcolor=black}
\usepackage{hyperref}

\begin{document}

\preprint{APS/123-QED}

\title{Magnon Planar Hall Effect and Anisotropic Magnetoresistance in a Magnetic Insulator}

\author{J. Liu}
\thanks{}
\email{jing.liu@rug.nl}
\affiliation{Physics of Nanodevices, Zernike Institute for Advanced Materials, University of Groningen, Groningen, The Netherlands}
\author{L. J. Cornelissen}
\affiliation{Physics of Nanodevices, Zernike Institute for Advanced Materials, University of Groningen, Groningen, The Netherlands}%

\author{J. Shan}
\affiliation{Physics of Nanodevices, Zernike Institute for Advanced Materials, University of Groningen, Groningen, The Netherlands}%
\author{T. Kuschel}
\affiliation{Physics of Nanodevices, Zernike Institute for Advanced Materials, University of Groningen, Groningen, The Netherlands}%
\author{B. J. van Wees}
\affiliation{Physics of Nanodevices, Zernike Institute for Advanced Materials, University of Groningen, Groningen, The Netherlands}

\date{\today}

\begin{abstract}
Electrical resistivities can be different for charge currents travelling parallel or perpendicular to the magnetization in magnetically ordered conductors or semiconductors, resulting in the well-known planar Hall effect and anisotropic magnetoresistance. Here, we study the analogous anisotropic magnetotransport behavior for magnons in a magnetic insulator Y$_{3}$Fe$_{5}$O$_{12}$. Electrical and thermal magnon injection, and electrical detection methods are used at room temperature with transverse and longitudinal geometries to measure the magnon planar Hall effect and anisotropic magnetoresistance, respectively. We observe that the relative difference between magnon current conductivities parallel and perpendicular to the magnetization, with respect to the average magnon conductivity, i.e. $|(\sigma_{\parallel}^{\textrm{m}}-\sigma_{\perp}^{\textrm{m}})/\sigma_{0}^{\textrm{m}}|$ , is approximately 5\% with the majority of the measured devices showing $\sigma_{\perp}^{\textrm{m}}>\sigma_{\parallel}^{\textrm{m}}$.

\end{abstract}


\maketitle

Different electrical resistivities for charge currents parallel and perpendicular to the magnetization were first discovered in ferromagnetic metals \cite{thomson1856electro}. Microscopically, it is understood as a second-order spin-orbit effect, which causes the anisotropic properties of the scattering between the conduction electrons and localized magnetic d-electrons \cite{kondo1962anomalous, ky1966plane, ky1967planar, kokado2012anisotropic}. These effects are applied in various technologies, for instance, magnetic recording and field sensoring \cite{livingston1996driving, freitas2007magnetoresistive}.

When a charge current with a current density of $j_{\textrm{x}}^{\textrm{c}}$ is applied parallel to the x-axis, electric fields perpendicular and parallel to $j_{\textrm{x}}^{\textrm{c}}$ build up as $E_{\textrm{xy}}^{\textrm{c}}$ and $E_{\textrm{xx}}^{\textrm{c}}$, depending on the angle $\alpha$ between $j_{\textrm{x}}^{\textrm{c}}$ and the in-plane magnetization. These can be described in a right-handed coordinate system as follows 
\begin{equation}
E_{\textrm{xy}}^{\textrm{c}}= j_{\textrm{x}}^{\textrm{c}} \, \Delta\rho^{\textrm{c}} \,\sin \alpha \,\cos \alpha,
\label{eqn:PHE}
\end{equation}
\begin{equation}
E_{\textrm{xx}}^{\textrm{c}}= j_{\textrm{x}}^{\textrm{c}}\, (\rho_{\perp}^{\textrm{c}}+\Delta\rho^{\textrm{c}} \,  \cos^{2}\, \alpha),
\label{eqn:AMR}
\end{equation}
with $\Delta\rho^{\textrm{c}}=\rho_{\parallel}^{\textrm{c}}-\rho_{\perp}^{\textrm{c}}$. $\rho_{\parallel}^{\textrm{c}}$ and $\rho_{\perp}^{\textrm{c}}$ are resistivities parallel and perpendicular to the magnetization. The planar Hall effect (PHE) is the transverse anisotropic magnetoresistance described by Eq.(\ref{eqn:PHE}), while the longitudinal anisotropic magnetoresistance captured in Eq.(\ref{eqn:AMR}) is denoted as AMR throughout this Letter. For most ferromagnetic metals, $\rho_{\parallel}^{\textrm{c}}>\rho_{\perp}^{\textrm{c}}$ \cite{mcguire1975anisotropic}. The magnitude of the effect, i.e. $\Delta\rho^{\textrm{c}}/\rho^{\textrm{c}}$, is in the order of 1$\%$. 

\begin{figure}[b]
	\includegraphics[width=0.99\linewidth]{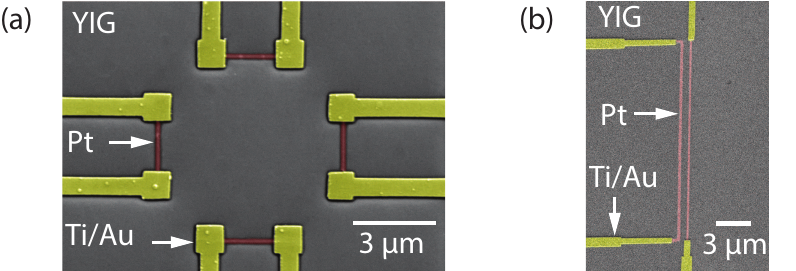}
	\caption{Colored scanning electron microscope (SEM) images of typical devices for (a) MPHE and (b) MAMR measurement. The yellow-colored structures are Ti/Au contacts and pink-colored ones are Pt strips. The grey background is the YIG substrate.}
	\label{fig:SEM}
\end{figure}

\begin{figure*}[t]
	\includegraphics[width=0.99\linewidth]{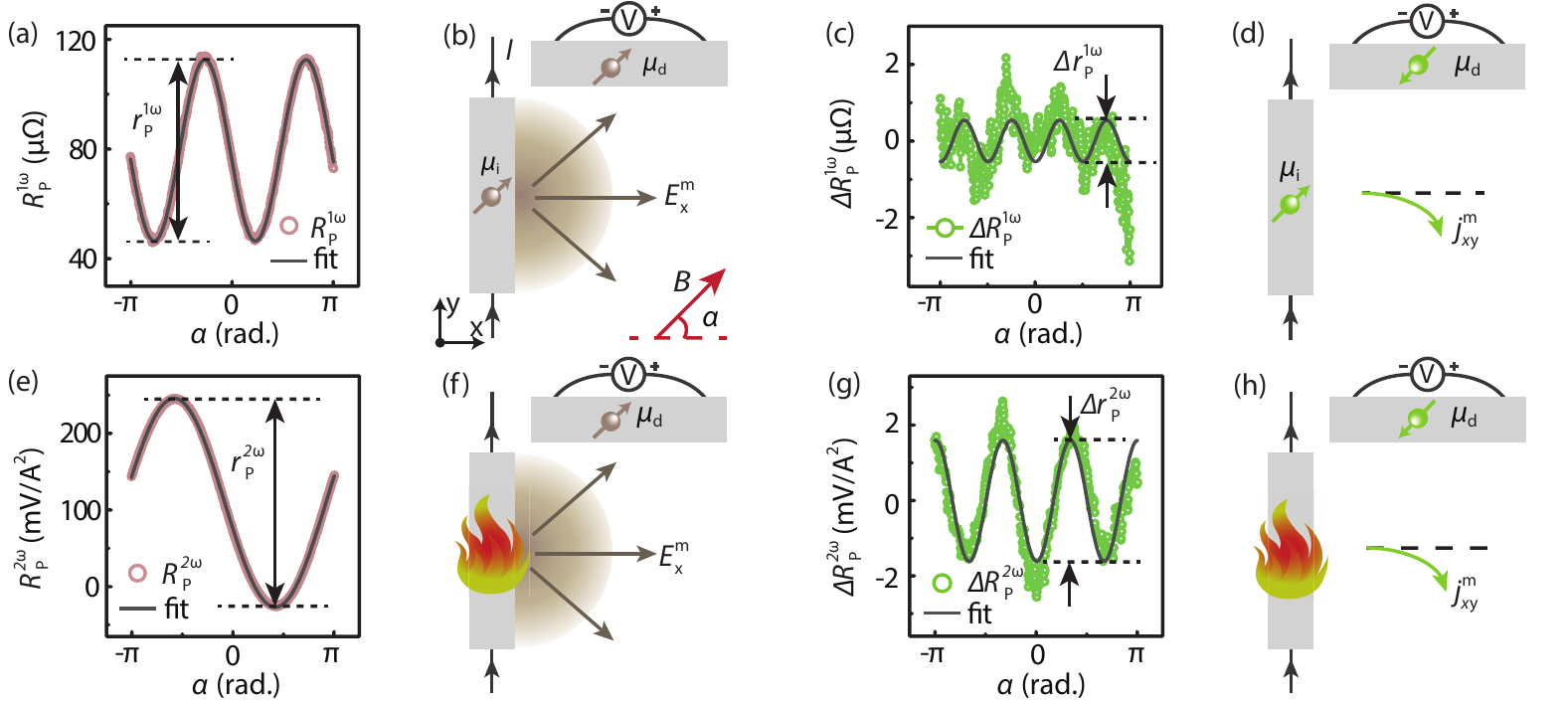}
	\caption{\label{fig:PHE_singledet} MPHE measurements for a typical device (series I, sample C, device 1). (a)-(d) First harmonic signal (electrical injection). (e)-(h) Second harmonic signal (thermal injection). (a), (b), (e), (f) Detection of the isotropic magnon current driven by the magnon chemical potential gradient, such as $E_{\textrm{x}}^{\textrm{m}}$. (c), (d), (g), (h) Detection of the MPHE current, $j_{\textrm{xy}}^{\textrm{m}}$. We perform a $\pi$ and $2\pi$ period sinusoidal fit for the measured $R^{1\omega}_{\textrm{P}}$ and $R^{2\omega}_{\textrm{P}}$ in (a) and (e). The residues of the fits are shown in (c) and (g) as $\Delta R^{1\omega}_{\textrm{P}}$ and $\Delta R^{2\omega}_{\textrm{P}}$, i.e. subtracting the $\pi$ and $2\pi$ period sinusoidal function from $R^{1\omega}_{\textrm{P}}$ and $R^{2\omega}_{\textrm{P}}$, respectively. Solid lines in (c) and (g) represent sinusoidal fits with period of $\pi/2$ and $2\pi/3$. The peak-to-peak amplitudes of the modulations are indicated as $r^{1\omega}_{\textrm{P}}$, $\Delta r^{1\omega}_{\textrm{P}}$, $r^{2\omega}_{\textrm{P}}$ and $\Delta r^{2\omega}_{\textrm{P}}$ in (a), (c), (e) and (g), respectively. (b), (d), (f), (h) Schematic illustration of a device top-view and measurement configuration. $\mu_{\textrm{i}}$ indicates the effective component of the magnon injection which is parallel to the magnetization aligned by $\textbf{\emph{B}}$ (40$\,$mT), while $\mu_{\textrm{d}}$ denotes the component sensored by the detector. In (b) and (f), the brown clouds represent isotropic magnon diffusion from the midpoint of the injector (in reality, the whole injector strip functions). In (f) and (h), the fire represents thermal injection of Joule heating from the electrical charge current.
	}
\end{figure*}

Magnons, or spin wave quanta, are the elementary excitations of magnetically ordered systems \cite{chumak2015magnon}. For long wavelength GHz spin waves, the dipolar interaction plays an important role, which is intrinsically anisotropic. This results in the anisotropic transport behavior for spin waves excited via microwave field \cite{kobljanskyj2015detection}. In contrast, for short wavelength THz spin waves, the Heisenberg exchange energy, i.e. $-J\, \textbf{S}_\textrm{i}\cdot \textbf{S}_\textrm{j}$, dominates the dispersion, resulting in isotropic magnon propagation. However, the asymmetric spin-orbital coupling, such as Dzyaloshinskii-Moriya interaction, can cause anisotropic transport of exchange magnons \cite{dzialoshinskii1957thermodynamic, PhysRev.120.91, PhysRevB.90.224403}.

Here, we report the observation of the PHE and AMR for magnon currents in a magnetic insulator at room temperature, the magnon planar Hall effect (MPHE) and magnon anisotropic magnetoresistance (MAMR), respectively. Magnons can carry both spins and heat. Since the 1960s, the thermal properties of magnetic insulators have been extensively studied to investigate spin wave transport \cite{akhiezer1958theory,bar1961scattering,LUTHI196235,PhysRev.121.1662,PhysRev.129.1132,PhysRev.122.388}. For example, Douglass \cite{PhysRev.129.1132} reported the anisotropic heat conductivities of the single crystal bulk ferrimagnetic insulator yttrium iron garnet (Y$_3$Fe$_5$O$_{12}$, YIG) with respect to the magnetic field at 0.5$\,$K. 

Recently, it has been reported that high energy exchange magnons ($E\sim k_{\textrm{B}}T$) can be excited thermally \cite{cornelissen2015long, giles2015long, PhysRevB.94.174437} and electrically \cite{cornelissen2015long, goennenwein2015non, PhysRevB.94.174405, ganzhorn2016magnon} and detected electrically in lateral non-local devices on YIG thin films. Later on, spin injection and detection in vertical sandwich devices was shown \cite{li2016observation, wu2016observation}. The magnon transport can be described by a diffusion-relaxation equation, with a characteristic magnon relaxation length of $\lambda_{\textrm{m}}\sim10 \, $$\mu$m for both electrically and thermally excited magnons at room temperature \cite{cornelissen2015long}. In this Letter our aim is to use this newly established electrical approach \cite{cornelissen2015long} to study the magnetotransport properties for exchange magnons in a magnetic insulator, where charge transport is prohibited due to the bandgap.

Typical devices used in our MPHE and MAMR measurements are shown in Fig.$\ $\ref{fig:SEM}. They are fabricated on single-crystal (111) YIG films with thickness of $100 \, $nm (series I) and $200\,$nm (series II). The saturation magnetization $M_{\textrm{s}}$ and Gilbert damping parameter $\alpha$ are comparable for the YIG samples in two series ($\mu_{0}M_{\textrm{s}}\sim170\,$mT, $\alpha\sim1\times 10^{-4}$). The YIG films are grown on a $500\, \mu $m thick (111) Gd$ _{3} $Ga$ _{5} $O$ _{12} $ (GGG) substrate by liquid-phase epitaxy and obtained commercially from Matesy GmbH. The Pt electrodes are defined using electron beam lithography followed by dc sputtering in Ar$^{+}$ plasma. The thickness of Pt layer is $\sim 7 \, $nm. The Ti/Au ($5/75 \, $nm) contacts are deposited by electron beam evaporation. Seven YIG samples are used with multiple devices on each of them. An overview of all devices is given in Supplementary Material VI \cite{liu2016MPHEMAMR_supplementarymaterial}.

Here, we use the electrical/thermal magnon excitation and electrical magnon detection method with Pt injectors/detectors on top of YIG as described in Ref.$\ $\cite{cornelissen2015long}. A low frequency ($ \omega/2\pi = 17.5\, $Hz) ac-current $I$ is sent through one Pt strip. It generates magnons in the YIG in two ways. First, the electrical current induces a transverse spin current due to spin Hall effect (SHE) \cite{PhysRevLett.83.1834, hoffmann2013spin}. This results in electron spin accumulation at the Pt$\mid$YIG interface, which can excite magnons in magnetic insulators via spin-flip scattering at the interface \cite{PhysRevLett.109.096603}. This is known as electrical magnon injection. Second, the Joule heating from the electrical current can thermally excite magnons via the bulk spin Seebeck effect \cite{PhysRevB.94.174437}. Other strips are used as magnon detectors, in which the spin current flowing into the detector is converted to a voltage signal due to the inverse spin Hall effect (ISHE) \cite{saitoh2006conversion}. Using lock-in technique, the electrically and thermally excited magnons can be measured as the first and second harmonic voltages separately. They scale linearly and quadratically with the current, i.e. $V^{1\omega}\sim I$ and $V^{2\omega}\sim I^{2}$, respectively (see Appendix A in Ref.$\ $\cite{vlietstra2014simultaneous}). Here, we normalize them by $I$ as non-local resistances ($R^{1\omega}=V^{1\omega}/I$ and $R^{2\omega}=V^{2\omega} /I^{2}$).

\begin{figure}[t]
	\includegraphics[width=0.99\linewidth]{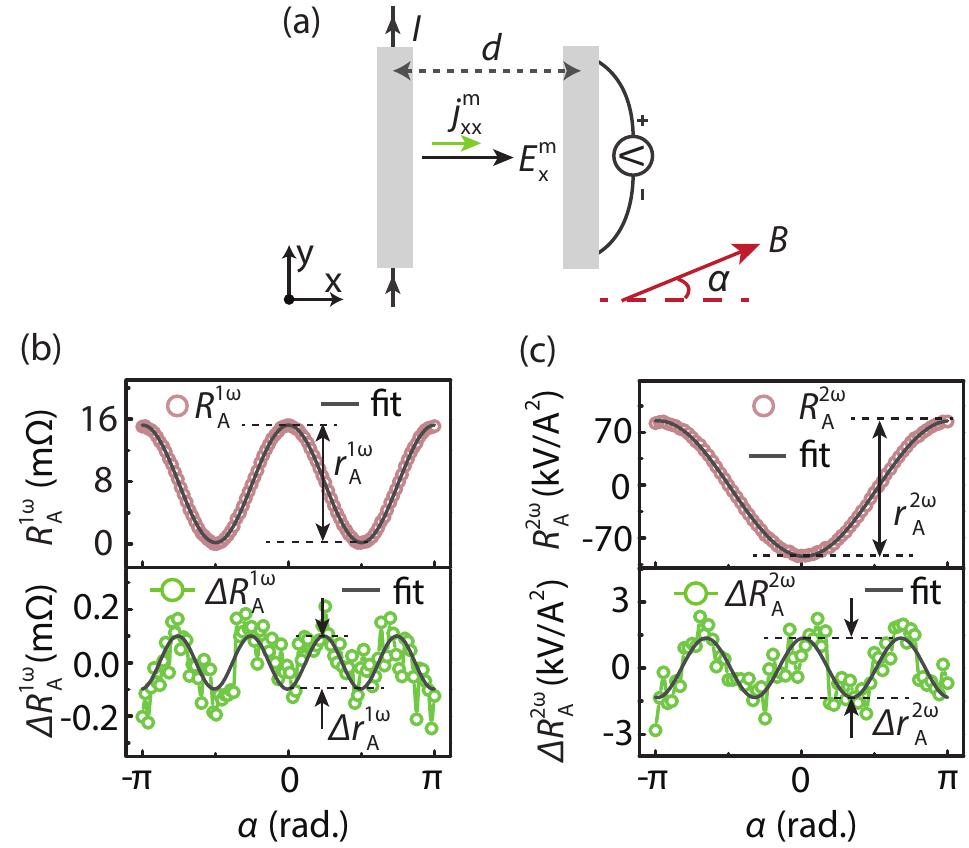}
	\caption{MAMR measurement for a typical device (series II, sample F, device 1). (a) Schematic top-view of the measurement configuration. The spacing between injector and detector is indicated as $d$. (b) First and (c) second harmonic signals with $d = 200 \, $nm, i.e. $R^{1\omega}_{\textrm{A}}$ and $R^{2\omega}_{\textrm{A}}$. The solid lines are $\pi$- and $2\pi$-period sinusoidal fits. In the lower panels of (b) and (c), the residues of the fits, i.e. the difference between data and corresponding fits, are shown as $\Delta R^{1\omega}_{\textrm{A}}$ and $\Delta R^{2\omega}_{\textrm{A}}$. They are fitted with $\pi/2$- and $2\pi/3$-period sinusoidal functions, respectively. $B=20\,$mT.}
	\label{fig:AMR}
\end{figure}

For the MPHE measurements, we use an injector and detector which are perpendicular to each other, while MAMR measurements employ a detector parallel to the injector. The magnon chemical potential gradient \cite{PhysRevB.94.014412}, which is created by the non-equilibrium magnons excited by the injector, drives the diffusion of the magnons in YIG. We define the direction which is perpendicular to the injector strip as the longitudinal direction with $E_{\textrm{x}}^{\textrm{m}}$ being the longitudinal magnon chemical potential gradient. We measure the transverse and longitudinal magnon currents with current densities of $j_{\textrm{xy}}^{\textrm{m}}$ and $j_{\textrm{xx}}^{\textrm{m}}$, i.e. the number of magnons passing through per unit cross-sectional area per second (see Figs.$\ $\ref{fig:PHE_singledet}(b), \ref{fig:PHE_singledet}(d)) and \ref{fig:AMR}(a)). 

Different from the PHE and AMR measurement for charge currents, we measure the magnetization direction dependent currents instead of the voltages. This is confirmed by the geometric reduction of the non-local signal by increasing the distance between Pt injector and detector on top of YIG within the diffusion regime for magnon transport \cite{cornelissen2015long}. Therefore, the nonlocal magnon transport measurement quantified by the non-local resistances detects the magnon conductivity $\sigma^{\textrm{m}}$ instead of the resistivity. However, in this Letter we still keep the terms, such as anisotropic magnetoresistance for MAMR, because of the analogous magnetotransport behaviors of electrons and magnons. 

An in-plane magnetic field $\textbf{\emph{B}}$ is applied to align the magnetization of the YIG film with an angle $\alpha$. We vary $\alpha$ by rotating the sample in-plane under a static magnetic field with a stepper motor. The MPHE and MAMR currents are expected to have angular dependences of
\begin{equation}
j_{\textrm{xy}}^{\textrm{m}}=E_{\textrm{x}}^{\textrm{m}} \, \Delta\sigma^{\textrm{m}} \,\sin\alpha\,\cos\alpha,
\label{eqn:MPHE_j}
\end{equation}
\begin{equation}
j_{\textrm{xx}}^{\textrm{m}}=E_{\textrm{x}}^{\textrm{m}} (\sigma_{\perp}^{\textrm{m}}+\Delta\sigma^{\textrm{m}}  \cos ^{2}\alpha),
\label{eqn:MAMR_j}
\end{equation}
where $\Delta \sigma^{\textrm{m}}=\sigma_{\parallel}^{\textrm{m}}-\sigma_{\perp}^{\textrm{m}}$. $\sigma_{\parallel}^{\textrm{m}}$ and $\sigma_{\perp}^{\textrm{m}}$ are conductivites for the magnon currents parallel and perpendicular to the magnetization direction, respectively.

The result of the first harmonic MPHE measurement for electrically injected magnons in Fig.$\ $\ref{fig:PHE_singledet}(a) shows mainly a $\pi$-period angular dependence. This is already discussed in prior works \cite{cornelissen2015long} and shown in Fig.$\ $\ref{fig:PHE_singledet}(b). A charge current is sent through the injector, by which a spin accumulation is created at the Pt$\mid$YIG interface via the SHE. The effective component for the magnon injection, i.e. $\mu_{\textrm{i}}$, is parallel to the magnetization. This results in a $\cos\alpha$ injection efficiency \cite{cornelissen2015long}. An isotropically diffusing magnon current propagates along the magnon chemical potential gradient \cite{PhysRevB.94.014412}, being directly detected as $\mu_{d}$. Due to the ISHE, a charge voltage is measured with an efficiency of $\sin\alpha$. Taking both injection and detection into account, we end up with a $\pi$-period sinusoidal modulation
\begin{equation}
R^{1\omega}_{\textrm{P}}\sim C^{1\omega}\sigma_{0}^{\textrm{m}}\cos\alpha\sin\alpha=\frac{1}{2}C^{1\omega}\sigma_{0}^{\textrm{m}}\sin2\alpha,
\label{ang_1st}
\end{equation} which corresponds to the angular dependence shown in Fig.$\ $\ref{fig:PHE_singledet}(a). $C^{1\omega}$ is a constant related to electrical magnon injection and detection efficiency and $\sigma_{0}^{\textrm{m}}$ is average magnon current conductivity. Details are explained in Supplementary Material II \cite{liu2016MPHEMAMR_supplementarymaterial} (including Ref. \cite{PhysRevB.93.020403}). 

For the residue of the $\pi$-period sinusoidal fit, i.e. the discrepancy between the data and fit (Fig.$\ $\ref{fig:PHE_singledet}(c)), there is a $\pi/2$-period sinusoidal modulation in the first harmonic signals. This is ascribed to the existence of the MPHE as illustrated in Fig.$\ $\ref{fig:PHE_singledet}(d). The MPHE induces an additional $\pi$-period angular dependence as indicated in Eq.$\ $(\ref{eqn:MPHE_j}). Together with the injection-detection efficiencies described in Eq. (\ref{ang_1st}), i.e. $(C^{1\omega}\cos\alpha \, \sin\alpha) \, (\Delta \sigma^{\textrm{m}} \sin \alpha\,\cos\alpha)$, it results in a component in the first harmonic resistance with an angular dependence of
\begin{equation}
\Delta R^{1\omega}_{\textrm{P}} \sim-\frac{1}{8}C^{1\omega}\Delta\sigma^{\textrm{m}}\cos 4\alpha.
\label{ang_MPHE_1st}
\end{equation}
This corresponds to the $\pi/2$-period modulation in Fig.$\ $\ref{fig:PHE_singledet}(c). 

For the second harmonic MPHE measurement, the thermal injection due to the Joule heating is insensitive to the YIG magnetization. Therefore, the thermally excited magnons can be directly detected as electron spins with polarization parallel to the magnetization as $\mu_{\textrm{d}}$ (c.f. Fig.$\ $\ref{fig:PHE_singledet}(f)) with a detection efficiency of $\sin \alpha$
\begin{equation}
R^{2\omega}_{\textrm{P}} \sim C^{2\omega} \sigma_{0}^{\textrm{m}}\sin\alpha,
\label{ang_2nd}
\end{equation}
which corresponds to the $2\pi$-period modulation in Fig.$\ $\ref{fig:PHE_singledet}(e). $C^{2\omega}$ is a parameter describing the thermal injection and electrical detection efficiency which is explained further in Supplementary Material II \cite{liu2016MPHEMAMR_supplementarymaterial}. Since the electrically and thermally excited magnons show a similar $\lambda_{\textrm{m}}$ over a wide temperature range \cite{cornelissen2015long, cornelissen2016temperature} and a similar magnetic field dependent behaviour \cite{PhysRevB.93.020403}, this strongly suggests that the same exchange magnons are involved in the spin transport. Therefore, we assign the same conductivities $\sigma_{0}^{\textrm{m}}$ and $\Delta \sigma^{\textrm{m}}$ to electrically and thermally excited magnons.  

Similarly, by looking at the deviation of the data from the $2\pi$-period modulation, a $2\pi/3$-period oscillation is observed in Fig.$\ $\ref{fig:PHE_singledet}(g).  When the thermal magnons also experience the MPHE, i.e. $(C^{2\omega}\,\sin\alpha) \, (\Delta \sigma^{\textrm{m}} \sin \alpha\,\cos\alpha)$, we expect a component in the second harmonic signal as 
\begin{equation}
\Delta R^{2\omega}_{\textrm{P}} \sim-\frac{1}{4}C^{2\omega}\Delta\sigma^{\textrm{m}}\cos 3\alpha,
\label{ang_MPHE_2nd}
\end{equation} 
which conforms to the $2\pi/3$-period oscillation in Fig.$\ $\ref{fig:PHE_singledet}(g). Besides, we also did MPHE measurement by using two detectors which are symmetrically patterned with respect to the injector, where we obtain the doubled asymmetric MPHE-current and the suppressed isotropic magnon current due to symmetry. Also, it excludes the influence of the asymmetric potential gradient in the single detector case (explained in detail in Supplementary Material III \cite{liu2016MPHEMAMR_supplementarymaterial}).

To quantify the MPHE, we extract the peak-to-peak amplitude of $R^{1\omega}_{\textrm{P}}$, $\Delta R^{1\omega}_{\textrm{P}}$, $R^{2\omega}_{\textrm{P}}$ and $\Delta R^{2\omega}_{\textrm{P}}$ as $r_{\textrm{P}}^{1\omega}$, $\Delta r_{\textrm{P}}^{1\omega}$, $r_{\textrm{P}}^{2\omega}$ and $\Delta r_{\textrm{P}}^{2\omega}$ by using 
\begin{equation}
R^{1\omega}_{\textrm{P}}=\frac{1}{2} r_{\textrm{P}}^{1 \omega}\, \sin(2\alpha+\alpha_{1})+R_{1} \label{R_1w_p},
\end{equation}
\begin{equation}
\Delta R^{1\omega}_{\textrm{P}}=-\frac{1}{2} \Delta   r_{\textrm{P}}^{1 \omega}\, \cos(4\alpha+\alpha_{2})+R_{2}, \label{D_R_1w_p}
\end{equation}
\begin{equation}
R^{2\omega}_{\textrm{P}}=\frac{1}{2} r_{\textrm{P}}^{2 \omega}\, \sin(\alpha+\alpha_{3})+R_3,
\label{R_2w_p}
\end{equation}
\begin{equation}
\Delta R^{2\omega}_{\textrm{P}}=-\frac{1}{2} \Delta   r_{\textrm{P}}^{2 \omega}\, \cos(3\alpha+\alpha_{4})+R_{4},
\label{D_R_2w_p}   
\end{equation}
with angle shifts indicated as $\alpha_{1}, \alpha_{2}, \alpha_{3}$ and $\alpha_{4}$, and offsets expressed as $R_{1}, R_{2}, R_{3}$ and $R_{4}$. They vary in different device geometries and measurement configuration. Further details are explained in Supplementary Material I \cite{liu2016MPHEMAMR_supplementarymaterial}. 

We obtain the magnitude of the MPHE as $\Delta \sigma^{\textrm{m}}/\sigma^{\textrm{m}}_{0}$ by determining $\Delta r^{\textrm{n}\omega}_{\textrm{P}}/r^{\textrm{n}\omega}_{\textrm{P}}$ according to approximate Eqs. (\ref{ang_1st})-(\ref{ang_MPHE_2nd}) and Eqs. (\ref{R_1w_p})-(\ref{D_R_2w_p})
\begin{equation}
\frac{\Delta \sigma^{\textrm{m}}}{\sigma^{\textrm{m}}_{0}}\approx\frac{4\,\Delta r^{1\omega}_{\textrm{P}}}{r^{1\omega}_{\textrm{P}}},
\label{eta_1}
\end{equation}
\begin{equation}
\frac{\Delta \sigma^{\textrm{m}}}{\sigma^{\textrm{m}}_{0}}\approx\frac{4\,\Delta r^{2\omega}_{\textrm{P}}}{r^{2\omega}_{\textrm{P}}},
\label{eta_2}
\end{equation} for the first and second harmonic signals, respectively. For the derivation, see Supplementary Materials II \cite{liu2016MPHEMAMR_supplementarymaterial}. For the results shown in Fig. \ref{fig:PHE_singledet}, we extract the magnitude of  $|\Delta \sigma^{\textrm{m}}/\sigma^{\textrm{m}}_{0}|$ as $(6.6\pm0.6)\, \%$ and $(4.7\pm0.2)\, \%$ for the first and second harmonic signals, respectively. Regarding the sign, we observe that $\Delta\sigma^{\textrm{m}}<0$, i.e. $\sigma_{\parallel}^{\textrm{m}}<\sigma_{\perp}^{\textrm{m}}$, for both first and second harmonic signals, since $r^{1\omega}_{\textrm{P}}$, $r^{2\omega}_{\textrm{P}}<0$ and $\Delta r^{1\omega}_{\textrm{P}}$, $\Delta r^{2\omega}_{\textrm{P}}>0$  in Fig. \ref{fig:PHE_singledet}. This sign agrees with the results of the heat conductivity measurement on the single crystal YIG at low temperature, when mainly magnons carry the heat \cite{PhysRev.129.1132}.
\begin{figure}[t!]
	\includegraphics[width=0.99\linewidth]{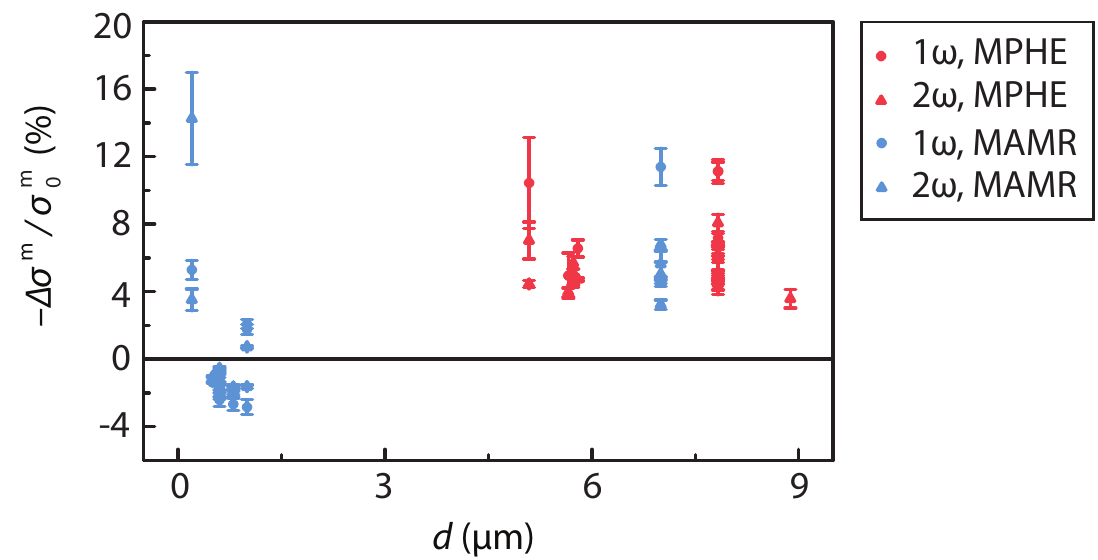}
	\caption {Sign and amplitude of the MPHE and MAMR measurements. $-\Delta\sigma^{\textrm{m}}/\sigma^{\textrm{m}}_{0}$ as a function of the injector-to-detector spacing $d$. Solid circles and triangles denote the first and second harmonic signals, i.e. $1\omega$ and $2\omega$, while pink and blue colors represent MPHE and MAMR results, respectively. $-\Delta\sigma^{\textrm{m}}>0$ means $\sigma_{\perp}^{\textrm{m}}>\sigma_{\parallel}^{\textrm{m}}$. The sign anomaly appears for the MAMR devices with $d\in [{0.2,1.0}]\,\mu$m. In this regime, the magnitude of $-\Delta\sigma^{\textrm{m}}/\sigma^{\textrm{m}}_{0}$ is comparably smaller. Without considering the data with the anomalous sign, we calculate the average value of $-\Delta\sigma^{\textrm{m}}/\sigma^{\textrm{m}}_{0}$ as $(6.1\pm2.1)\%$ and $(5.0\pm4.0)\%$ for the MPHE and MAMR, respectively. For the MPHE device, $d$ is defined as the spacing between the middle points of the injector and detector.}
	\label{sign_mag_MPHE_MAMR}
\end{figure}

In the MAMR measurements, we also observe the characteristic period for the first and second harmonic signals, a $\pi/2$-period and a $2\pi/3$-period angular modulation, respectively (see Figs.$\ $\ref{fig:AMR}(b) and \ref{fig:AMR}(c)). For the magnitude of the MAMR results, we can extract the peak-to-peak amplitudes of  $R^{1\omega}_{\textrm{A}}$, $\Delta R^{1\omega}_{\textrm{A}}$, $R^{2\omega}_{\textrm{A}}$, $\Delta R^{2\omega}_{\textrm{A}}$ as $r^{1\omega}_{\textrm{A}}$, $\Delta r^{1\omega}_{\textrm{A}}$, $r^{2\omega}_{\textrm{A}}$, $\Delta r^{2\omega}_{\textrm{A}}$ from the results shown in Fig.$\,$\ref{fig:AMR}. We obtain $|\Delta\sigma^{\textrm{m}}/\sigma^{\textrm{m}}_{0}|$ as $(5.3\pm0.6) \,\%$ and $(5.9\pm0.6) \,\%$ for the first and second harmonic signals with the same sign of $\sigma_{\parallel}^{\textrm{m}}<\sigma_{\perp}^{\textrm{m}}$. 

The sign and magnitude of all the measured MPHE and MAMR are summarized in Fig. \ref{sign_mag_MPHE_MAMR}. On different samples and devices, all the MPHE devices show the sign of $\sigma_{\perp}^{\textrm{m}}>\sigma_{\parallel}^{\textrm{m}}$ for both first and second harmonic signals. However, as can be seen in Fig. \ref{sign_mag_MPHE_MAMR}, for the MAMR measurement, the oppsite sign and weaker effect arises when the injector-to-detector spacing is in the range of $[{0.2,1.0}]\,\mu$m. We do not understand, why the sign and magnitude anomaly appears in this range. More details are described in Supplementary Material VI \cite{liu2016MPHEMAMR_supplementarymaterial}.

We exclude possible extra modulations induced by the misalignment between the magnetic field and in-plane magnetization angle due to the anisotropy or sample misalignment as described in Supplementary Materials IV and V \cite{liu2016MPHEMAMR_supplementarymaterial} (including Refs. \cite{wang2014strain,meyer2016observation,spaldin2010magnetic,ya1986growth}). Besides, we check the reciprocity and linearity for $R^{1\omega}$ and $\Delta R^{1\omega}$ in Supplementary Materials VII \cite{liu2016MPHEMAMR_supplementarymaterial}.

To conclude, we observe MPHE and MAMR for both electrically and thermally injected magnons from the angular dependent transverse and longitudinal non-local measurement at room temperature. The magnitude of these effects, $|\Delta\sigma^{\textrm{m}}/\sigma_{0}^{\textrm{m}}|$, is approximately 5\% for both electrically and thermally injected magnons on YIG thin films, which is in the same order of magnitude as that of PHE or AMR in ferromagnetic metals. We observe that $\sigma_{\perp}^{\textrm{m}}>\sigma_{\parallel}^{\textrm{m}}$ for all the measured devices except those MAMR devices with certain injector-to-detector spacing. This is similar to the electronic magnetoresistance of most metallic systems ($\rho_{\parallel}^{\textrm{c}}>\rho_{\perp}^{\textrm{c}}$). Our results establish a new way to study and employ the magnetotransport of magnons, which can give an insight into the spin-orbital interaction of insulating materials. 
 
We acknowledge H. M. de Roosz, J. G. Holstein, H. Adema and T. J. Schouten for their technical
assistance and appreciate M. Mostovoy, T.T.M Palstra, I. J. Vera Marun, J. Mendil and R. Schlitz for discussion. This work is part of the research programme of the Foundation for Fundamental Research on Matter (FOM) and supported by NanoLab NL and the Zernike Institute for Advanced Materials. Further support by EU FP7 ICT Grant No. 612759 InSpin and the Deutsche Forschungsgemeinschaft (DFG) within the priority program Spin Caloric Transport (SPP 1538, KU3271/1-1) is gratefully acknowledged.

\bibliography{reference}

\end{document}


\renewcommand{\thesection}{S.\arabic{section}}
\renewcommand{\thesubsection}{\thesection.\arabic{subsection}}
 
%
\makeatletter 
\makeatother

\subsection {\Large \uppercase{Supplemental Material}}
\subsection{\Large Magnon Planar Hall Effect and Anisotropic Magnetoresistance in a Magnetic Insulator}
\begin{center}
	\large J. Liu*, L.J. Cornelissen, J. Shan, T. Kuschel, B.J. van Wees
\end{center}
\subsection{}
In this supplemental material, we discuss the origin of the angle shift in the angular dependent MPHE and MAMR measurement (section $\textcolor{blue}{\textrm{I}}$) and the derivation of the formulas to calculate the magnitude of the MPHE and MAMR (section $\textcolor{blue}{\textrm{II}}$). Moreover, the result of a double-detector MPHE measurement is shown in section $\textcolor{blue}{\textrm{III}}$. Besides, we exclude other possible additional angular modulations caused by any possible misalignment between magnetization and magnetic field, including the in-plane magnetocrystalline anisotropy (section $\textcolor{blue}{\textrm{IV}}$) and the out-of-plane tilt of the sample plane with respect to the applied magnetic field (section $\textcolor{blue}{\textrm{V}}$). Then, we give the summary of the sign and amplitude of the MPHE and MAMR on different samples and devices (section $\textcolor{blue}{\textrm{VI}}$). Last, we verify that the linearity and Onsager reciprocity hold for the MAMR (section $\textcolor{blue}{\textrm{VII}}$).

\subsection{\large I. Origin of the angle shift in the MPHE and MAMR measurement}

\begin{figure}[b!]
	\includegraphics[width=0.6\linewidth]{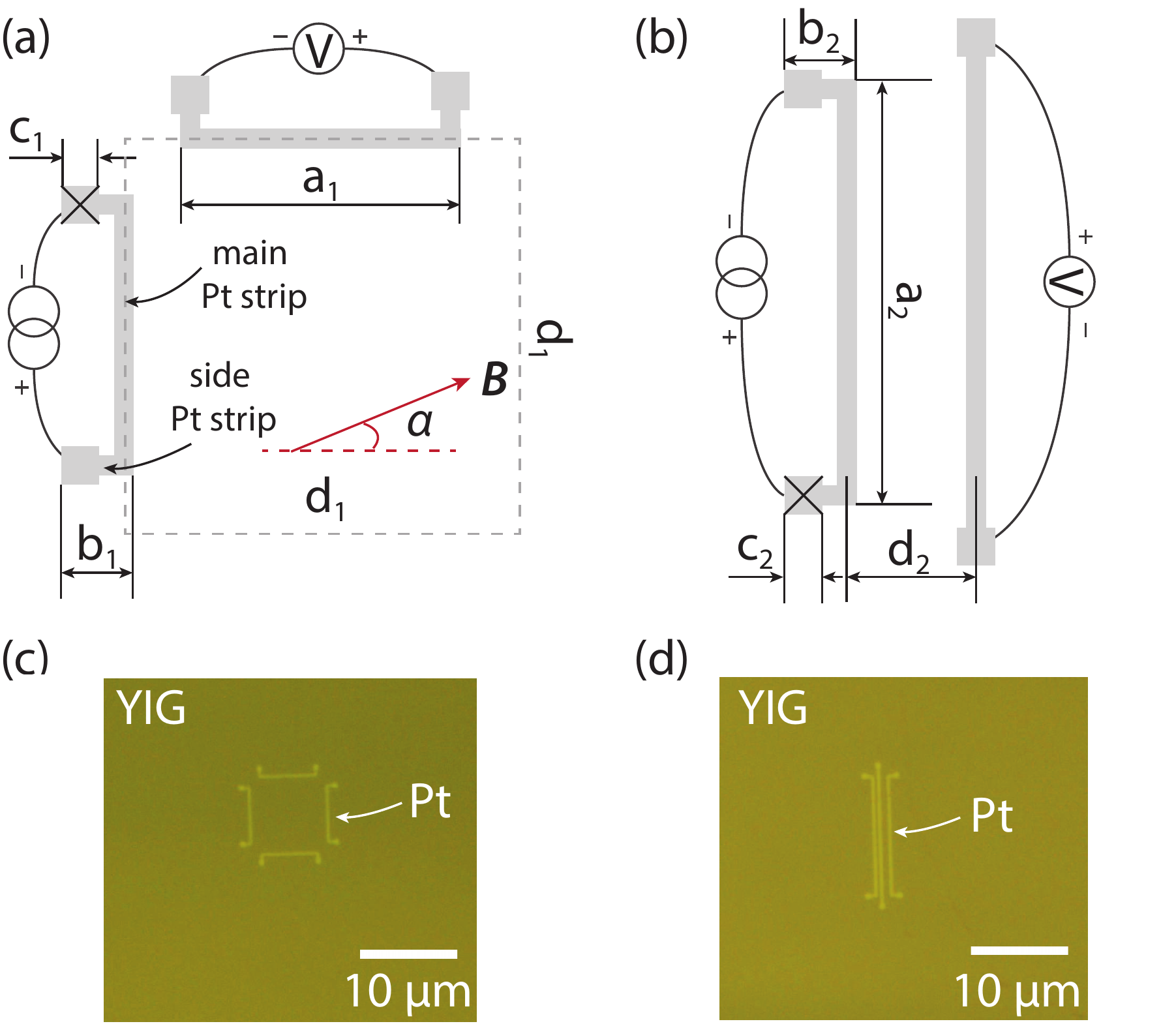}
	\caption{Schematic illustration of top-views for the design of Pt strips for (a) MPHE and (b) MAMR devices. The main strips have lengths of $a_{1}$ and $a_{2}$ while the side strips are with lengths of $b_{1}$ and $b_{2}$. In (a), the square with side length of $d_{1}$ in dashed grey line is a reference. The injector and detector Pt strips are designed symmetrically with respect to the center of this reference square with the same structure. Black crosses denote square shapes with side lengths of $c_{1}$ and $c_{2}$. Corresponding optical images for typical (c) MPHE and (d) MAMR device before depositing Ti/Au electrodes.}
	\label{figS1}
\end{figure}

The angle $\alpha$ is defined such that $\alpha=0$ when the magnetic field is perpendicular to the injector Pt strip as shown in Fig. \ref{figS1}. Ideally, for example, in Fig. \ref{figS1}(a), the magnon detection depends on $-\sin \alpha$, which is the angular dependence we expect for the second harmonic signals. The minus sign is due to the polarity of the voltmeter. However, we observe a ($-\sin \alpha$)-modulation with an angle shift of $\alpha_{3}$ in the angular dependence of $R^{2\omega}_{\textrm{P}}$ as shown in Fig. $\textcolor{blue}{2}$(e) and captured in Eq. $(\textcolor{blue}{11})$. Similarly, a small angle shift from $-\sin2\alpha$ oscillation is observed for $R^{1\omega}_{\textrm{P}}$ in Fig. $\textcolor{blue}{2}$(a). This is mainly due to the following reasons.

Firstly, in experiment it is hard to precisely control the alignment of the devices with respect to the magnetic field. Therefore, $\alpha$ has a small error bar, approximately $\pm\,5$ degrees. 

Secondly, in the design of devices, for the convenience of making connection between Pt and Ti/Au electrodes, side strips are designed at the end of the Pt main strips as shown in Figs.$\ $\ref{figS1}(a) and (b). These structures are not visible in SEM images of the typical devices for MPHE and MAMR measurement shown in Fig. $\textcolor{blue}{1}$, since they are covered by the Ti/Au electrodes. However, they can be seen in the optical images of the devices before depositing Ti/Au electrodes (Figs.$\ $\ref{figS1}(c) and (d)). They also function as an injector or detector but with a 90 degree rotation angle compared to the main Pt strips with length of $a_1$ and $a_2$ in Figs.$\ $\ref{figS1}(a) and (b), respectively. The influence of these side strips is less pronounced in MAMR measurements than that in the MPHE measurements. This is because in MAMR measurements the contribution of the side strips at two ends of the injector or detector cancel out due to the symmetry, while this is not the case in MPHE measurements. For example, in the second harmonic MPHE measurements, one side strip of the detector is closer to the heater than the one at the other end. Since the device dimensions are smaller than the magnon spin diffusion length, the signals decrease geometrically with increasing the distance between injector and detector \cite{cornelissen2015long}. The signals picked up by the side strips at the two ends of detector due to ISHE do not cancel out. This gives rise to a detection contribution from the side strip closer to the injector with an efficiency of $\cos \alpha$. It results in an angle shift from the $\sin \alpha$-modulation that we expect from the main Pt detector strip. The magnitude of this angle shift in MPHE measurements depends on the relative contribution of the main and side strips. The magnitude of the signal also scales with the length of the device \cite{cornelissen2015long}. Since the length of the main strip is larger than that of the side strip ($a_1\colon b_{1}\approx6$), the contribution of the main strip in this aspect is larger. However, the average distance from the side strip to the heater is smaller than that for the main strip. Therefore, in terms of the spacing between injector to detector, the detection of the side strip is more efficient than that of the main strip. We summarize the angle shifts in Eqs. ($\textcolor{blue}{9}$)-($\textcolor{blue}{12}$) as $\alpha_{1}$, $\alpha_{2}$, $\alpha_{3}$ and $\alpha_{4}$.

Here, we give the qualitative explanation above in order to show that the misalignment of angle $\alpha$ and the influence of the side strip do not produce other oscillation periods for the angular dependence of $R^{1\omega}$ and $R^{2\omega}$ but only cause a small angle shift. This does not affect our determination of the MPHE and MAMR based on their periodic characteristics in the magnetic field angle sweeping measurements.

\subsection{\large II. Derivation for the magnitude of MPHE and MAMR signals}
\begin{figure}[b!]
	\includegraphics[width=0.5\linewidth]{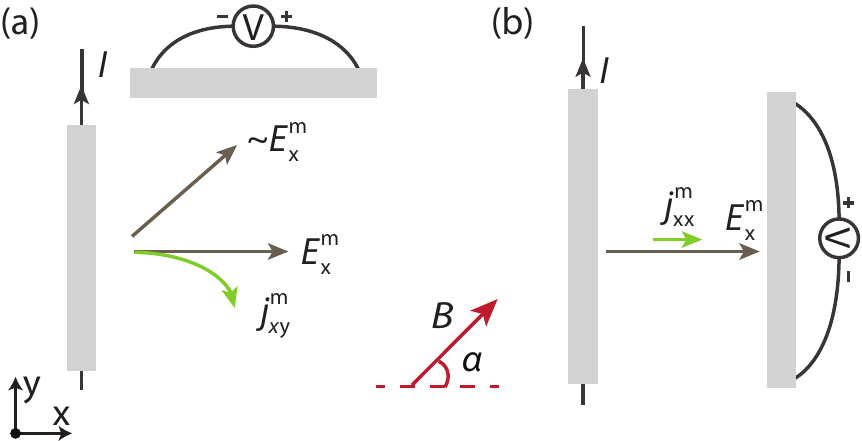}
	\caption{Schematic illustration of top-views for (a) MPHE and (b) MAMR devices in an xy-coordinate system. $E_{\textrm{x}}^{\textrm{m}}$ is the longitudinal magnon chemical potential gradient along the x-axis created by the injector where a current of $I$ is sent. In (a), the magnon chemical potential gradient between the injector and detector is approximately $E_{\textrm{x}}^{\textrm{m}}$. A magnetic field $B$ is applied with an angle $\alpha$ to control the in-plane magnetization direction. $j_{\textrm{xy}}^{\textrm{m}}$ and  $j_{\textrm{xx}}^{\textrm{m}}$ are the transverse and longitudinal magnon current densities which depend on the in-plane magnetization direction as described by Eqs. ($\textcolor{blue}{3}$) and ($\textcolor{blue}{4}$). }
	\label{S_fig_derivation}
\end{figure}
Analogous to the electron transport of conducting system in the diffusive regime where electrons move along the electrochemical potential gradient with a certain electrical conductivity, magnons diffuse in the magnetic insulator driven by the gradient of the magnon chemical potential with a magnon conductivity \cite{PhysRevB.94.014412}. In our MPHE and MAMR devices, the distances between the injector and detector are smaller than the characteristic magnon spin diffusion length $\lambda_{\textrm{m}}$ at room temperature \cite{cornelissen2015long}, so that the magnon transport we discuss here is in the diffusive regime. By using this theory, we derive the angular dependence of the MPHE and MAMR measurements, i.e. $R^{\textrm{n}\omega}_{\textrm{X}}$ and $\Delta R^{\textrm{n}\omega}_{\textrm{X}}$ ($n=1$ or 2: first or second harmonic signals; $\textrm{X}=\textrm{P}$ or A: MPHE or MAMR results), as shown in Eqs. ($\textcolor{blue}{5})\textrm{-}(\textcolor{blue}{8})$. Based on this, we obtain the expression for the defined magnitude of the MPHE and MAMR, i.e. $\Delta\sigma^{\textrm{m}}/\sigma^{\textrm{m}}_{0}$ (see Eqs. $(\textcolor{blue}{13})$ and $(\textcolor{blue}{14})$), in terms of the measurement results, i.e. the magnitude and sign of angular oscillation for $R^{\textrm{n}\omega}_{\textrm{X}}$ and $\Delta R^{\textrm{n}\omega}_{\textrm{X}}$ ($r^{\textrm{n}\omega}_{\textrm{X}}$ and $\Delta r^{\textrm{n}\omega}_{\textrm{X}}$) in Figs. $\textcolor{blue}{2}$ and $\textcolor{blue}{3}$ and Eqs. $(\textcolor{blue}{9})\textrm{-}(\textcolor{blue}{12})$.

In Fig. \ref{S_fig_derivation}, we define a longitudinal magnon chemical potential gradient $E_{\textrm{x}}^{\textrm{m}}$, which is perpendicular to the injector strip. In MAMR measurement, we can directly probe this $E_{\textrm{x}}^{\textrm{m}}$. In contrast, for the MPHE measurement in Fig. \ref{S_fig_derivation}(a) where injector and detector are perpendicular to each other, the magnon chemical potential gradient between the injector and detector can be different. However, here we assume it is approximately $E_{\textrm{x}}^{\textrm{m}}$, because we consider the edge effect of the injector is small. Moreover, $E_{\textrm{x}}^{\textrm{m}}$ created by the SHE-induced magnon spin accumulation or the thermal gradient from the Joule heating of the current is the driving force for the diffusion of the magnons. They can be detected separately as the electrically and thermally excited magnons by the first and second harmonic signals; therefore, we denote them as $E_{\textrm{x}}^{\textrm{m}}(\textrm{n}\omega)$,
\begin{equation}
E_{\textrm{x}}^{\textrm{m}}(\textrm{n}\omega)=C_{\textrm{i}}^{\textrm{n}\omega}\,I^{\textrm{n}}\,(\cos \alpha)^{2-n} \label{E_0_1w}
\end{equation}
where $C_{\textrm{i}}^{\textrm{1}\omega}$ [VA$^{-1}$m$^{-1}$] or $C_{\textrm{i}}^{\textrm{2}\omega}$ [VA$^{-2}$m$^{-1}$] are the electrical and thermal magnon injection factors, in which the subscript "i" represents injection. They describe the conversion efficiencies from the electrical charge current ($I$) or corresponding Joule heating ($\sim I^{2}$) to the magnon chemical potential gradients, respectively. Electrical magnon injection depends on the in-plane magnetization angle $\cos \alpha$ in both measurement geometries as shown in Figs. \ref{S_fig_derivation}(a) and (b), while the thermal injection is independent of the in-plane magnetization direction. 

Considering the isotropically propagating magnons, the built-up magnon potential gradient in a certain direction leads to a magnon diffusion current in the same direction $j^{\textrm{m}}=\sigma^{\textrm{m}}_{0}\,E^{\textrm{m}}$, where $\sigma^{\textrm{m}}_{0}$ is the average magnon conductivity. We assign the same conductivities $\sigma_{0}^{\textrm{m}}$ and $\Delta \sigma^{\textrm{m}}$ to electrically and thermally excited magnons, since the same exchange magnons are involved in the spin transport. The magnon current is picked up by the detector strip and converted to a charge voltage $V^{\textrm{n}\omega}$ by the ISHE. We normalize $V^{\textrm{n}\omega}$ by the current $I^{\textrm{n}}$ as the non-local resistance $R^{\textrm{n}\omega}$. Therefore, in MPHE and MAMR shown in Figs. \ref{S_fig_derivation}(a) and (b) the magnon currents which diffuse directly from the injector to detector are measured as
\begin{equation}
R^{\textrm{n}\omega}_{\textrm{P}}\sim\frac{[\sigma^{\textrm{m}}_{0}\,E^{\textrm{m}}_{\textrm{x}}(\textrm{n}\omega)](C_{\textrm{d}}\sin\alpha)}{I^{\textrm{n}}}
\label{R_nw_MPHE}
\end{equation}

\begin{equation}
R^{\textrm{n}\omega}_{\textrm{A}}=\frac{[\sigma^{\textrm{m}}_{0}\,E^{\textrm{m}}_{\textrm{x}}(\textrm{n}\omega)](C_{\textrm{d}}\cos\alpha)}{I^{\textrm{n}}} \label{R_nw_MAMR}
\end{equation}
where $C_{\textrm{d}}$ [Vm$^{2}$A$^{-1}$] is a parameter  describing the conversion efficiency between the magnon current and ISHE-based charge voltage, where "d" stands for detection. $C_{\textrm{d}}$ is the same for electrically and thermally excited magnons. Thus, Eqs. (\ref{R_nw_MPHE}) and (\ref{R_nw_MAMR}) describe the direct isotropic transport of the electrically and thermally excited magnons in two geometries shown in Figs. \ref{S_fig_derivation}(a) and (b), respectively. We list the expression for the angular dependence oscillations of $R^{\textrm{n}\omega}_{\textrm{X}}$ in Table \ref{tab:derivation_sum}.

For the MPHE and MAMR, due to the difference between conductivities for the magnon currents parallel and perpendicular to the magnetization ($\sigma^{\textrm{m}}_{\parallel}\neq\sigma^{\textrm{m}}_{\perp}$), a transverse and longitudinal magnon current with current densities of $j_{\textrm{xy}}^{\textrm{m}}$ and $j_{\textrm{xx}}^{\textrm{m}}$ is generated with a driving force of the longitudinal magnon potential gradient $E^{\textrm{m}}_{\textrm{x}}$ according to Eqs. $(\textcolor{blue}{3})$ and $(\textcolor{blue}{4})$. They are also measured by the detector based on the ISHE and normalized by the current as
\begin{equation}
\Delta R^{\textrm{n}\omega}_{\textrm{P}}=\frac{j_{\textrm{xy}}^{\textrm{m}}(C_{\textrm{d}}\sin\alpha)}{I^{\textrm{n}}}=C^{\textrm{n}\omega}_{\textrm{i}}C_{\textrm{d}}(\Delta\sigma^{\textrm{m}}\sin\alpha\,\cos\alpha)(\cos\alpha)^{\textrm{2-n}}\,\sin\alpha
\label{D_R_nw_MPHE}
\end{equation}

\begin{equation}
\Delta R^{\textrm{n}\omega}_{\textrm{A}}=\frac{j_{\textrm{xx}}^{\textrm{m}}(C_{\textrm{d}}\cos\alpha)}{I^{\textrm{n}}}=C^{\textrm{n}\omega}_{\textrm{i}}C_{\textrm{d}}(\sigma^{\textrm{m}}_{\perp}+\Delta\sigma^{\textrm{m}}\cos^{2} \alpha)(\cos\alpha)^{\textrm{3-n}} \label{D_R_nw_MAMR}
\end{equation}
where the anisotropic magnetotransport properties are captured. We list the angular dependence oscillations of $\Delta R^{\textrm{n}\omega}_{\textrm{X}}$ only with the characteristic periods, i.e. $\pi/2$ for the first and $2\pi/3$ for the second harmonic signals, in Table \ref{tab:derivation_sum}.

Therefore, we obtain the expression for the magnitude of the MPHE and MAMR in Table \ref{tab:derivation_sum} , i.e. $\Delta\sigma^{\textrm{m}}/\sigma^{\textrm{m}}_{0}$, in terms of the measurement results, i.e. $\Delta r^{\textrm{n}\omega}_{\textrm{X}}/r^{\textrm{n}\omega}_{\textrm{X}}$, as shown in Eqs. $(\textcolor{blue}{13})$ and $(\textcolor{blue}{14})$. From the relative sign and magnitude of $\Delta r^{\textrm{n}\omega}_{\textrm{X}}$ and $r^{\textrm{n}\omega}_{\textrm{X}}$, we can determine the sign of $\Delta \sigma^{\textrm{m}}=\sigma^{\textrm{m}}_{\parallel}-\sigma^{\textrm{m}}_{\perp}$ and the magnitude of $\Delta\sigma^{\textrm{m}}/\sigma^{\textrm{m}}_{0}$. Here, we neglect the contribution to the $\pi/2-$ and $\pi$-period oscillations caused by the multiplication of the sinusoidal dependent function due to MPHE and MAMR as shown in Eqs. (\ref{D_R_nw_MPHE}) and (\ref{D_R_nw_MAMR}), because their amplitudes are negligibly small compared with those of $R^{\textrm{n}\omega}_{\textrm{X}}$ in Eqs. (\ref{R_nw_MPHE}) and (\ref{R_nw_MAMR}).

{\renewcommand{\arraystretch}{1.6}%

\begin{table}[h!]
	\caption{\label{tab:derivation_sum}Summary of the expression for $R^{\textrm{n}\omega}_{\textrm{X}}$, $\Delta R^{\textrm{n}\omega}_{\textrm{X}}$ and $\Delta\sigma^{\textrm{m}}/\sigma^{\textrm{m}}_{0}$ for MPHE and MAMR measurements.}
	\begin{ruledtabular}
		\begin{tabular}{ccccc}
			type of measurement&\multicolumn{2}{c}{MPHE ($\textrm{X}=$P)} &\multicolumn{2}{c}{MAMR ($\textrm{X}=$A)}\\
			$\textrm{n}^{\textrm{th}}$ harmonic&1$\omega $ (n=1) &2$\omega$ (n=2)&1$\omega$  (n=1)&2$\omega$ (n=2)\\
			\hline
			$R^{\textrm{n}\omega}_{\textrm{X}}$&$\frac{1}{2}r^{1\omega}_{\textrm{P}}\sin 2\alpha$
			
			&$\frac{1}{2}r^{2\omega}_{\textrm{P}}\sin \alpha$
			
			& $\frac{1}{2}r^{1\omega}_{\textrm{A}}\cos 2\alpha$
			
			& $\frac{1}{2}r^{2\omega}_{\textrm{A}}\cos\alpha$ \\
			&($r^{1\omega}_{\textrm{P}}\sim C^{1\omega}\sigma^{\textrm{m}}_{0}$)
			
			&($r^{2\omega}_{\textrm{P}}\sim 2C^{2\omega}\sigma^{\textrm{m}}_{0}$)
			
			&($r^{1\omega}_{\textrm{A}}=C^{1\omega}\sigma^{\textrm{m}}_{0}$)
			
			&($r^{2\omega}_{\textrm{A}}=2C^{2\omega}\sigma^{\textrm{m}}_{0}$)\\
			\hline
			$\Delta R^{\textrm{n}\omega}_{\textrm{X}}$&$-\frac{1}{2}\Delta r^{1\omega}_{\textrm{P}}\cos4\alpha$
			
			&$-\frac{1}{2}\Delta r^{2\omega}_{\textrm{P}}\cos3\alpha$
			
			& $\frac{1}{2}\Delta r^{1\omega}_{\textrm{A}}\cos4\alpha$
			
			& $\frac{1}{2}\Delta r^{1\omega}_{\textrm{P}}\cos3\alpha$ \\
			
			&($\Delta r^{1\omega}_{\textrm{P}}=\frac{1}{4}C^{1\omega}\Delta\sigma^{\textrm{m}}$)
			
			&($\Delta r^{2\omega}_{\textrm{P}}=\frac{1}{2}C^{2\omega}\Delta\sigma^{\textrm{m}}$)
			
			&($\Delta r^{1\omega}_{\textrm{A}}=\frac{1}{4}C^{1\omega}\Delta\sigma^{\textrm{m}}$)
			
			&($\Delta r^{2\omega}_{\textrm{A}}=\frac{1}{2}C^{2\omega}\Delta\sigma^{\textrm{m}}$)\\
			\hline
			$\Delta\sigma^{\textrm{m}}/\sigma^{\textrm{m}}_{0}$	&$\sim\frac{4\,\Delta r^{1\omega}_{\textrm{P}}}{r^{1\omega}_{\textrm{P}}} $
			&$\sim\frac{4\,\Delta r^{2\omega}_{\textrm{P}}}{r^{2\omega}_{\textrm{P}}} $& $\sim\frac{4\,\Delta r^{1\omega}_{\textrm{A}}}{r^{1\omega}_{\textrm{A}}} $& $\sim\frac{4\,\Delta r^{2\omega}_{\textrm{A}}}{r^{2\omega}_{\textrm{A}}} $		
		\end{tabular}
	\end{ruledtabular}
	\footnotetext[1]{$C^{\textrm{n}\omega}=C^{\textrm{n}\omega}_{\textrm{i}}C_{\textrm{d}}$}
\end{table}

\subsection{\large III. Double detector MPHE measurements}

\begin{figure}[b!]
	\includegraphics[width=0.7\linewidth]{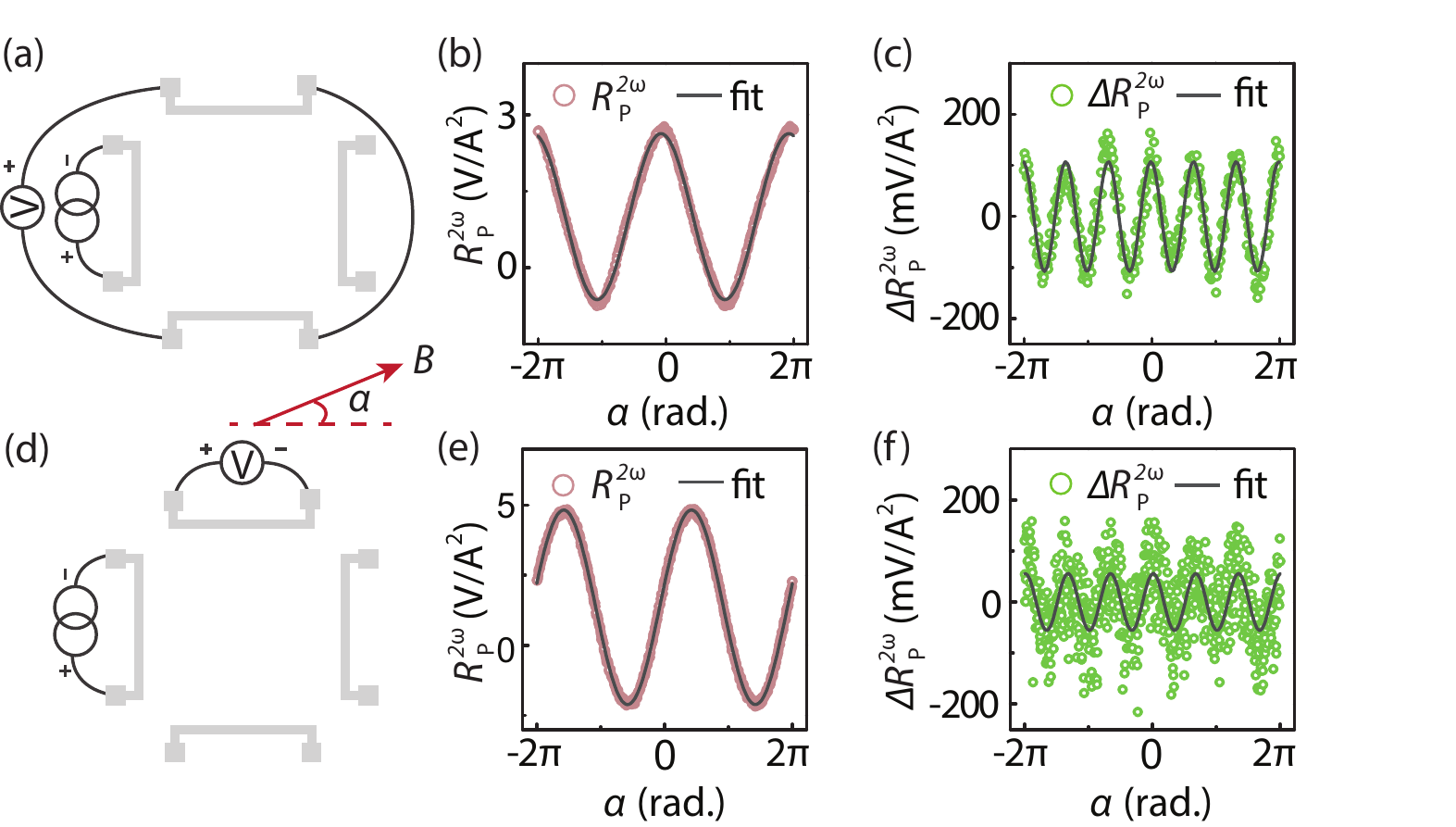}
	\caption{Comparison of the double and single detector MPHE measurement. (a), (d) Schematic illustration of measurement configuration with corresponding results for (b), (e) $R^{2\omega}_{\textrm{P}}$  and (c), (f) $\Delta R^{2\omega}_{\textrm{P}}$.}
	\label{figS4}
\end{figure}

In this section, we discuss an extra experiment we did for the MPHE, by which we can measure $\Delta R^{2\omega}_{\textrm{P}}$ with approximately double magnitude while $R^{2\omega}_{\textrm{P}}$ is suppressed, compared with the results shown in Fig. $\textcolor{blue}{(2)}$. Besides, this also confirms that the transverse magnon current we measured is not due to an asymmetric magnon potential gradient caused by the single-detector MPHE measurement configuration. 

As shown in Fig. \ref{figS4}(a), we used two detectors which are patterned symmetrically with respect to the injector strip. Compared with the single detector case in Fig. \ref{figS4}(d), $R^{2\omega}_{\textrm{P}}$ is reduced (compare Figs.$\ $\ref{figS4}(b) and (e)). This is because the isotropic magnon signals measured by the upper and lower main detector strip with the same polarization cancel out due to symmetry when we connect two detectors in such a way. The component of $R^{2\omega}_{\textrm{P}}$ which is left is mainly due to the side strip. By contrast, $\Delta R^{2\omega}_{\textrm{P}}$ is approximately doubled compared with that in single-detector measurement (compare Figs.$\ $\ref{figS4}(c) and (f)). This results from the fact that the MPHE-induced transverse magnon currents measured by the upper and lower detectors with different polarization add up in this measurement configuration. It also confirms that the MPHE-induced $2\pi/3$-period sinusoidal modulations do not result from parasitic effects with amplitudes scaling with the amplitude of the $2\pi$-period sinusoidal modulation for $R^{2\omega}_{\textrm{P}}$.

\subsection{\large IV. In-plane magnetocrystalline anisotropy of YIG (111)}
In this section, we investigate the possible influence of the in-plane magnetocrystalline anisotropy on the in-plane angle dependent measurement, in order to differentiate it from the features caused by the MPHE and MAMR.

\begin{figure}[t!]
	\includegraphics[width=0.99\linewidth]{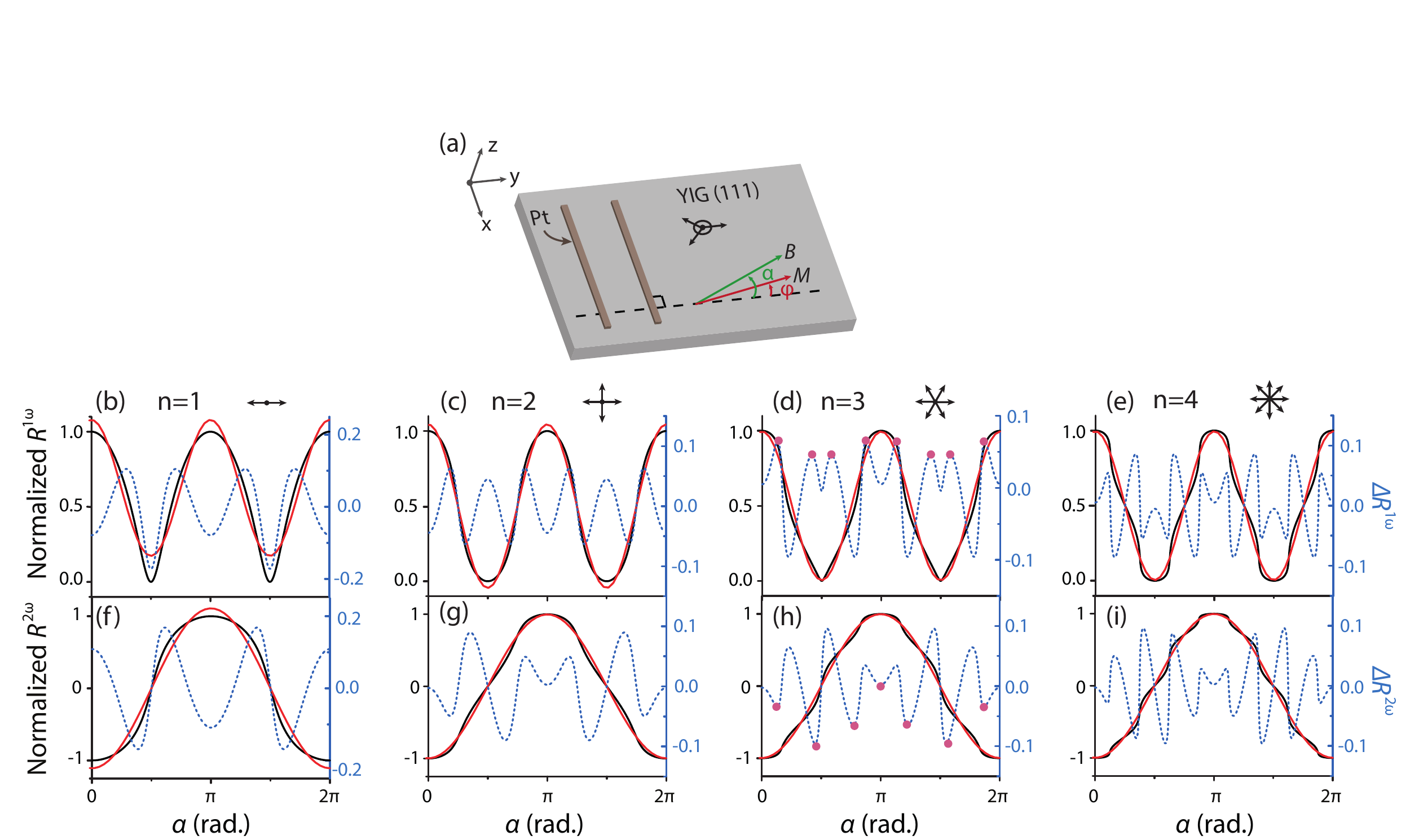}
	\caption {Simulation of the first and second harmonic signals under the influence of in-plane magnetocrystalline anisotropy. (a) Schematic illustration of the Pt injector and detector on top of YIG (111) thin film. The black dashed line is a reference line perpendicular to the injector and detector strips. In-plane magnetic field $\textbf{\textit{B}}$ is applied with an angle of $\alpha$ with respect to the reference line, while the magnetization of the YIG film $\textbf{\textit{M}}$ has an angle of $\varphi$. Three black arrow lines on top of YIG (111) represent the projection of three crystallographic directions, i.e. [100], [010] and [001]. Normalized (b)-(e) first harmonic signals, i.e. $R^{1\omega}$, and (f)-(i) second harmonic signals, i.e. $R^{2\omega}$, are shown as black solid lines with different number of magnetic easy axes ($n=1,2,3,4$). The black arrow lines represent $2n$ number of magnetocrystalline easy directions, along which the YIG film is more easily magnetized in the plane of the film compared with other in-plane directions. Red solid lines are $\pi$-period sinusoidal fits in (b)-(e) and  2$\pi$-period sinusoidal fits in (f)-(i). Blue dashed lines are the residues of the fits, i.e. the difference between the black and red solid lines, denoted as $\Delta R^{1\omega}$ and $\Delta R^{2\omega}$. The maximal value points of $\Delta R^{1\omega}$ in (d) and minimal value points of $\Delta R^{2\omega}$ in (h) are marked as the pink dots.}
	\label{fig_S5_model}
\end{figure}

We used YIG films with thickness of $100\,$nm and $200\,$nm. Due to the shape anisotropy, the magnetization prefers to align in the plane of the film. The in-plane saturation field is smaller than $1\,$mT, while the out-of-plane saturation field is more than $200\,$mT. Therefore, the YIG film roughly has a magnetic hard axis, i.e. perpendicular to the film surface, and a magnetic easy plane, i.e. the surface plane. Within this magnetic easy plane, it is easier to magnetize YIG along some crystallographic directions than the others. This magnetocrystalline anisotropy is intrinsically caused by the spin-orbit coupling and the coupling between the orbital and lattice \cite{spaldin2010magnetic}. Extrinsically, the strain from crystal growth can also modify the magnetocrystalline anisotropy \cite{wang2014strain}. In our experiment, we always use a YIG sample with (111) surface determination prepared by liquid-phase epitaxy (LPE) method. As a garnet structure, YIG has a cubic crystal structure. YIG (111) surface has a 3-fold rotation symmetry, i.e. C$_{3}$. 

Considering this in-plane magnetocrystalline anisotropy by applying the Stoner-Wohlfarth model (see Fig. \ref{fig_S5_model}(a)), we can write the magnetism-related energy density $\varepsilon_{\textrm{m}}$ as
\begin{equation}
\varepsilon_{\textrm{m}}=K\,\sin^{2}(n\,\varphi)-B\,M_{\textrm{s}}\cos(\alpha-\varphi)
\end{equation}
where $\alpha$ and $\varphi$ are the in-plane rotation angles of the magnetic field and magnetization, $K$ is the anisotropy constant, $B$ is the external magnetic field and $M_{\textrm{s}}$ is the saturation magnetization. $n$ is the number of magnetic easy axes, which means the system is more easily magnetized in $2n$ directions in the plane of the film. Here, we treat these $2n$ directions equally. Also, for simplicity, we assume that one of the magnetic easy axes is perpendicular to the detector strip.

In order to minimize the energy density $\varepsilon_{\textrm{m}}$, we let the derivative of $\varepsilon_{\textrm{m}}$ in terms of $\varphi$ equal zero ($\partial \varepsilon_{\textrm{m}}/\partial \varphi =0$). After confirming this is a minimal point, we can obtain the relation between $\alpha$ and $\varphi$ as
\begin{equation}
n\,K\,\sin(2n\,\varphi)=B\,M_{\textrm{s}}\sin(\alpha-\varphi)
\label{angle_relation}
\end{equation}
which describes how much the sweeping angle of the magnetization $\varphi$ deviates from the rotation angle of the external field $\alpha$, depending on the anisotropy constant $K$ and the strength of the external field $\textbf{\textit{B}}$. 

For the non-local measurements, we always control the angle $\alpha$ of the external field by using the rotating sample holder in a static magnetic field. However, the magnon injection and detection efficiency depend on the angle of the magnetization $\varphi$. In the device geometry as shown in Fig.$\ $\ref{fig_S5_model}(a), first and second harmonic signals depend on $\varphi$ as a function of $\cos^{2}\varphi$ and $\cos \varphi$, respectively. Combined with the relation between $\alpha$ and $\varphi$ in Eq.$\ $(\ref{angle_relation}), we know the dependences of first and second harmonic signals as a function of $\alpha$ shown as black solid lines in Figs.$\ $\ref{fig_S5_model}(b)-(i). Both first and second harmonic signals change shapes by varying the number of magnetic easy axes, for example $n=1,2,3,4$ as shown in Fig. \ref{fig_S5_model}. We fit the normalized $R^{1\omega}$ and $R^{2\omega}$ by $\pi$- and $2\pi$-period sinusoidal functions and look at the residues of the fits, i.e. $\Delta R^{1\omega}$ and $\Delta R^{2\omega}$, as shown in the blue dashed lines. $BM_{\textrm{s}}/nK$ is in the order of 10 in the simulation results in Fig.$\ $\ref{fig_S5_model}.

\begin{figure}[t!]
	\includegraphics[width=0.45\linewidth]{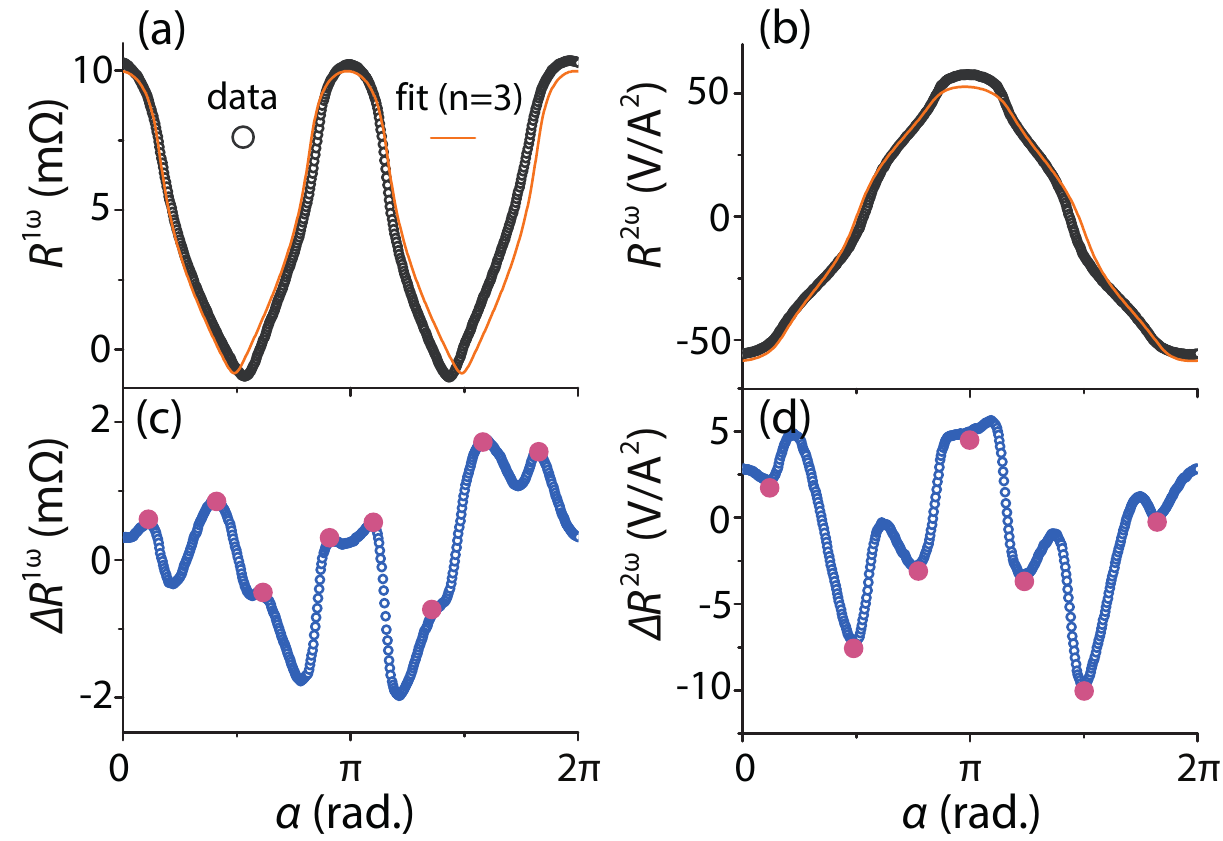}
	\caption {Experimental results of the small magnetic field measurement ($B=0.6\,$mT). (a) First  and (b) second harmonic signals with corresponding fits taking into account the in-plane magnetocrystalline anisotropy with three magnetic easy axes. (c) and (d) Residues of the $\pi$-period sinusoidal fit for the first harmonic signals and $2\pi$-period sinusoidal fit for the second harmonic signals. The maximal value points in (c) and minimal value points in (d) are marked as the pink dots.}
	\label{fig_S5_experiment}
\end{figure}

To check the influence of the in-plane crystallographic anisotropy experimentally, we applied a small field with magnitude of $0.6\,$mT to do the in-plane field angle dependent measurement. With such a small field, the in-plane angle of the YIG magnetization lags behind the external magnetic field angle. The extent of this lagging-behind behavior is modulated by the in-plane magnetocrystalline anisotropy of YIG (111) thin film. The results are shown in Fig.$\ $\ref{fig_S5_experiment}. Comparing it with the simulation results in Fig.$\ $\ref{fig_S5_model}, we find that the shape of the first and second harmonic signals (Figs.$\ $\ref{fig_S5_experiment}(a) and (b)) only conform to those in the case of three magnetic easy axes $(n=3)$ as shown in Figs.$\ $\ref{fig_S5_model}(d) and (h). We fit the measured $R^{1\omega}$ and $R^{2\omega}$ by the $(n=3)$-modified $\pi$- and $2\pi$-period sinusoidal functions in Figs.$\ $\ref{fig_S5_model}(d) and (h), respectively. The resulting symmetry of the magnetocrystalline anisotropy also agrees with the  crystallographic symmetry of YIG (111) plane, i.e. a 3-fold rotational symmetry (C$_3$). For the fitting parameters of both first and second harmonic signals, $BM_{\textrm{s}}/nK$ is in the order of 10. Based on this, we can estimate that $K\approx 0.3\times10^{3}\,$erg$\,$cm$^{-3}$, by taking $B=0.6\,$mT, $\mu_{0}M_{\textrm{s}}=170\,$mT and $n=3$. According to Ref. \cite{ya1986growth}, LPE-grown YIG films on the GGG substrate with tens of $\mu$m thickness show the anisotropy constant of $2.3\times10^{3}\,$erg$\,$cm$^{-3}$, which is comparable with our result.

Furthermore, we checked the residue of the $\pi$- and $2\pi$-period sinusoidal fits for the measured first and second harmonic signals and compared them with the simulation results. For the first harmonic signals, eight maximal value points as denoted as the pink points are observed in the residues for both simulation (Fig.$\ $\ref{fig_S5_model}(d)) and experimental results (Fig.$\ $\ref{fig_S5_experiment}(c)). In contrast, there are seven minimal value points for the second harmonic signals (see Figs.$\ $\ref{fig_S5_model}(h) and$\ $\ref{fig_S5_experiment}(d)). However, the shape of the residues in the simulation and experimental results are not exactly the same. This can be caused by the following reasons. First one is our assumption that one of the magnetic easy axes is perpendicular to the detector strip. However, in our device we do not know the exact relative orientation of the Pt strips with respect to the crystallographic orientation. Second, the assumption that the YIG can be equally easily magnetized along $2n$ directions might be not precisely true, especially with the two opposite directions along the same axis. Third, some other modulations also influence the shape of the first and second harmonic signals, for example, the $\pi$-period sinusoidal modulation in the second harmonic signals due to the spin Nernst effect in platinum \cite{meyer2016observation}. 

Here, we confirm that the characteristic features of the MPHE and MAMR, i.e. the $\pi/2$- and $2\pi/3$-period sinusoidal modulations for the first and second harmonic signals, are different from the modulation induced by the in-plane magnetocrystalline anisotropy, i.e. 8 maximum value points for the first harmonic signals and 7 minimum value points for the second harmonic signals within the $2\pi$ magnetic field angle sweeping. Moreover, we applied a magnetic field of $40\,$mT for the MPHE measurement as shown in Fig. $\textcolor{blue}{2}$ and $20\,$mT for the MAMR measurement as shown in Fig. $\textcolor{blue}{3}$. This means that $BM_{\textrm{s}}/nK$ is approximately in the order of 500. In Fig. \ref{fig_S5_quantitative}, we show the simulation results of the first and second harmonic signals with $BM_{\textrm{s}}/nK$ being 10 and 500 for $n=3$, respectively. They correspond to the small magnetic field measurement shown in Fig. \ref{fig_S5_experiment} and the MPHE (MAMR) measurement in Fig. $\textcolor{blue}{2}$ ($\textcolor{blue}{3}$). In Fig. \ref{fig_S5_experiment}, we see that the additional modulations, i.e. $\Delta R^{1\omega}$ or $\Delta R^{2\omega}$, induced by the in-plane magnetocrystalline anisotropy are about 10\% of $R^{1\omega}$ or $R^{2\omega}$. This is consistent with the simulation results in Figs. \ref{fig_S5_quantitative}(a) and (c). Then provided $BM_{\textrm{s}}/nK=500$, we can estimate the amplitude of the additional modulation caused by the in-plane magnetocrystalline anisotropy in the MPHE (MAMR) measurement configurations as shown in Figs. \ref{fig_S5_quantitative}(b) and (d). The amplitudes of the additional modulation, i.e. $\Delta R^{1\omega}$ or $\Delta R^{2\omega}$, are expected to be less than 0.2\% of the amplitudes of $R^{1\omega}$ or $R^{2\omega}$ in our MPHE and MAMR measurement. This is generally smaller than the amplitude of the MPHE (MAMR) signals, i.e. $\Delta r^{\textrm{n}\omega}_{\textrm{X}}/r^{\textrm{n}\omega}_{\textrm{X}}$, which is more than 1\%, corresponding to $\Delta\sigma^{\textrm{m}}/\sigma^{\textrm{m}}_{0}\sim5\%$ . Therefore, considering our qualitative analysis and quantitative estimation, we conclude that the observed modulations in our MPHE and MAMR measurements cannot be attributed to the magnetocrystalline anisotropy.

\begin{figure}[h!]
	\includegraphics[width=0.45\linewidth]{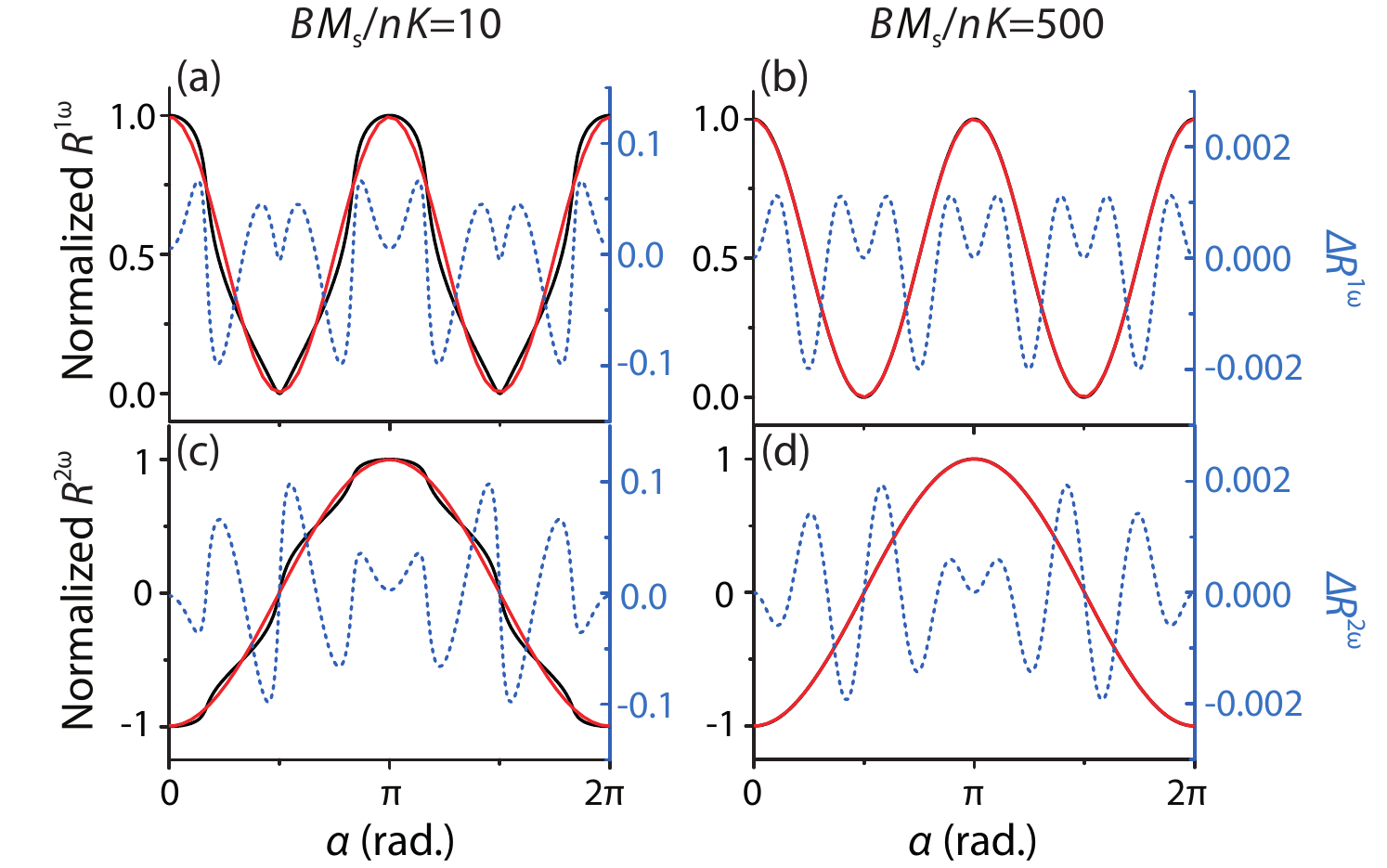}
	\caption {Influence of the in-plane magnetocrystalline anisotropy on the first and second harmonic signals under different strength of magnetic field. Normalized $R^{1\omega}$ and $R^{2\omega}$ with $BM_{\textrm{s}}/nK$ (a), (c) being 10 and (b), (d) being 500 corresponding to $B\sim0.6\,$mT and $30\,$mT, respectively ($K$=$0.3\times10^{3}\,$erg$\,$cm$^{-3}$, $\mu_{0}M_{\textrm{s}}\sim170\,$mT, $n=3$). Red curves are $\pi/2$-period sinusoidal fits in (a) and (b) and $\pi$-period sinusoidal fits in (c) and (d). Blue dashed lines are residues of the fits, denoting as $\Delta R^{1\omega}$ and $\Delta R^{2\omega}$ for the first and second harmonic signals, respectively.}
	\label{fig_S5_quantitative}
\end{figure}

\subsection{\large V. Out-of-plane misalignment of the sample plane with respect to the applied magnetic field}\label{Section_S6}
In this section, we quantitatively study the influence of the out-of-plane sample misalignment on the angle dependent measurement, in order to confirm that the characteristic $\pi/2$- and $2\pi/3$-period sinusoidal modulations for the first and second harmonic signals are due to the MPHE and MAMR. 

\begin{figure}[h!]
	\includegraphics[width=0.7\linewidth]{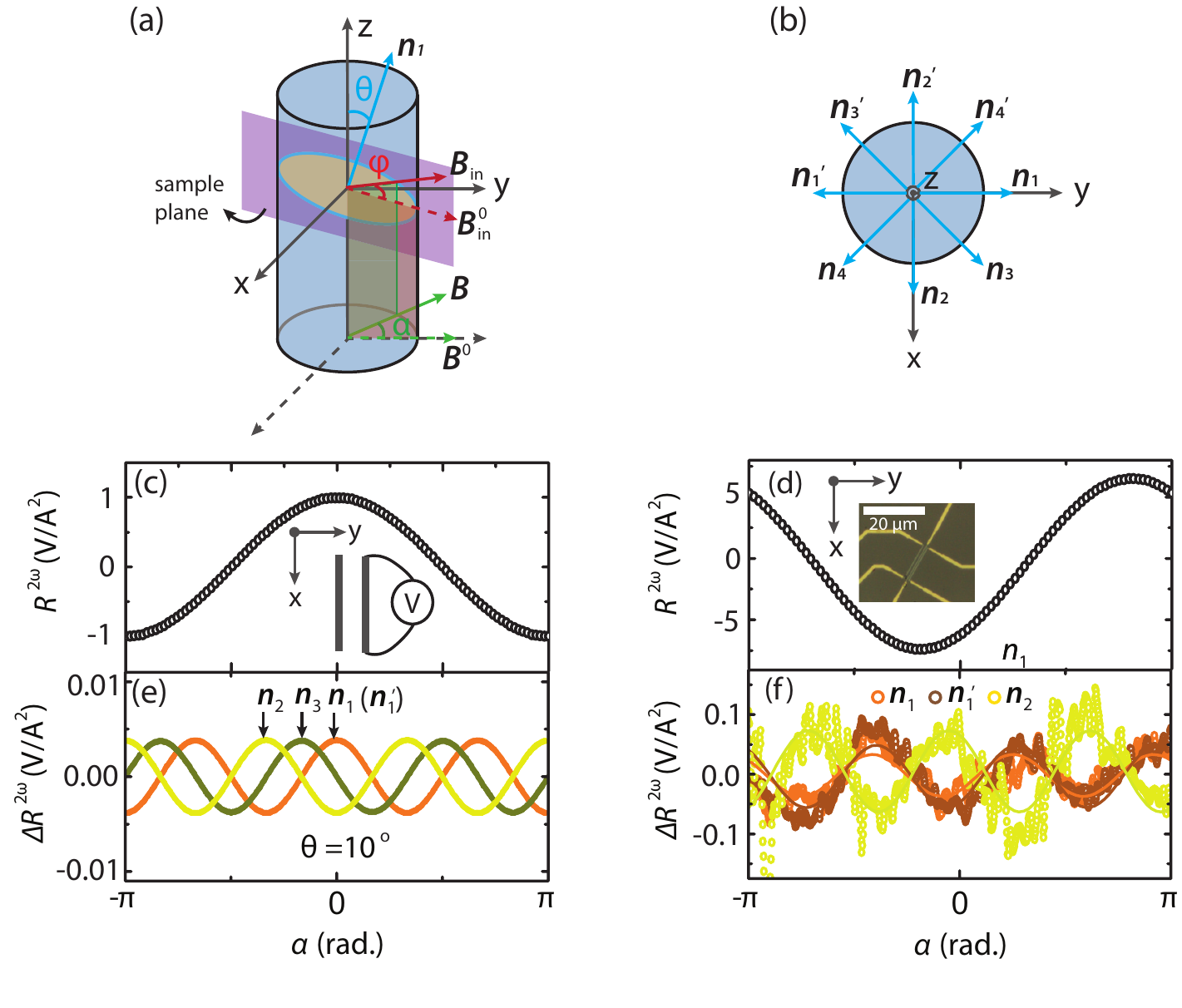}
	\caption {(a) Schematic illustration of the tilted sample plane with respect to the external magnetic field in a Cartesian coordinate system. The blue cylinder is for the convenience of visualization. An applied magnetic field $\textbf{\textit{B}}$ (green arrow line) rotates in xy-plane with an angle $\alpha$ (green) with respect to the positive y-axis. The purple plane represents the sample plane with a cross intersection with the cylinder. For example, the sample is tilted in a way that the angle between its normal vector $\textbf{\textit{n}}_{1}$ (blue arrow line) and the positive z-axis is $\theta$ (blue). The angle between the rotating in-plane projection of $\textbf{\textit{B}}$, i.e. $\textbf{\textit{B}}_{\textrm{in}}$, and the initial in-plane projection of $\textbf{\textit{B}}^{0}$, i.e. $\textbf{\textit{B}}^{0}_{\textrm{in}}$,  is $\varphi$ (red). (b) Topview of the different sample normal vector projections on the xy-plane. Qualitative comparison between (c), (e) the calculated  and (d), (f) experimental (series II, sample E, device 20 with $B=40\,$mT) results of the angular dependence of the second harmonic signals. The $2\pi$-period sinusoidal component of the second harmonic signals due to the detection efficiency according to (c) simulation (normalized) and (d) experimental results with topviews of device orientation as shown in the inset, respectively. (e) The simulation results of the $2\pi/3$-period sinusoidal components due to sample tilting out-of-plane with $\theta=10\,^{\textrm{o}}$ in different tilting directions. (f) The measured $2\pi/3$-period sinusoidal components with sample plane normal vectors of $\textbf{\textit{n}}_1$, $\textbf{\textit{n}}_{1}^{\prime}$ and $\textbf{\textit{n}}_{2}$.}
	\label{fig_S6}
\end{figure}

In our experimental setup, the sample is mounted on a rotating sample holder with rotation motor under a static magnetic field. We load the sample with its surface as parallel as possible with respect to the center-to-center line between the two magnetic poles according to the scale of the sample holder with accuracy of $\pm2\,^{\textrm{o}}$. This uncertainty can result in a sweeping angle $\varphi$ of the in-plane magnetization being different from the rotation angle of the external magnetic field $\alpha$. 

Here, we simplify the scenario as shown in Figs.$\,$\ref{fig_S6}(a) and (b). In the coordinate system defined in Fig. \ref{fig_S6}(a), the applied magnetic field $\textbf{\textit{B}}$ can be expressed as 
\begin{equation}
\textbf{\textit{B}} =(-B\sin\alpha, B\cos\alpha, 0).
\label{Eq_fieldB}
\end{equation}
For simplicity, we assume that the sample is static while the magnetic field $\textbf{\textit{B}}$ rotates with angle $\alpha$. The normal vector of the sample plane $\textbf{\textit{n}}$ can have a tilting angle $\theta$ with respect to the positive z-axis. Here, we assume the simple case that $\textbf{\textit{n}}$ is in yz-plane, i.e. $\textbf{\textit{n}}_1$, as shown in Fig. \ref{fig_S6}(a) and expressed as
\begin{equation}
\textbf{\textit{n}}=(0,\sin\theta,\cos\theta).
\label{Eq_n}
\end{equation}
 We can decompose the magnetic field $\textbf{\textit{B}}$ into two components, perpendicular and parallel to the sample plane denoted as $\textbf{\textit{B}}_{\textrm{out}}$ and $\textbf{\textit{B}}_{\textrm{in}}$ 
 \begin{equation}
 \textbf{\textit{B}}_{\textrm{out}}=\mid\textbf{\textit{B}}\mid\,\frac{\textbf{\textit{B}}\cdot\textbf{\textit{n}}}{\mid\textbf{\textit{B}}\mid\,\mid\textbf{\textit{n}}\mid}\,\textbf{\textit{n}}=(0,B\cos\alpha\sin^{2}\theta,B\cos\alpha\sin\theta\cos\theta)
 \label{Eq_Bout}
 \end{equation}
 \begin{equation}
 \textbf{\textit{B}}_{\textrm{in}}=\textbf{\textit{B}}-\textbf{\textit{B}}_{\textrm{out}}=(-B\sin\alpha,B\cos\alpha\cos^{2}\theta,-B\cos\alpha\sin\theta\cos\theta).
 \label{Eq_Bin}
 \end{equation}
 
  We assume that the initial position of the magnetic field is along the positive y-axis, so that the initial in-plane magnetic field $\textbf{\textit{B}}^{0}_{\textrm{in}}$ is expressed as 
  \begin{equation}
 \textbf{\textit{B}}_{\textrm{in}}^{0}=\textbf{\textit{B}}_{\textrm{in}}(\alpha=0)=(0,B\cos^{2}\theta,-B\sin\theta\cos\theta).
  \label{Eq_Bin0}
  \end{equation} 
Then, we can obtain the sweeping angle of the in-plane magnetic field $\varphi$, i.e. the angle between $\textbf{\textit{B}}^{0}_{\textrm{in}}$ and $\textbf{\textit{B}}_{\textrm{in}}$. We end up with 
\begin{equation}
\cos\varphi=\frac{\textbf{\textit{B}}_{\textrm{in}}\cdot\textbf{\textit{B}}_{\textrm{in}}^{0}}{\mid\textbf{\textit{B}}_{\textrm{in}}\mid\mid\textbf{\textit{B}}_{\textrm{in}}^{0}\mid}=\frac{\cos\theta\cos\alpha}{\sqrt{\cos^{2}\theta\cos^{2}\alpha+\sin^{2}\alpha}}
\label{Eq_cosvarphi}
\end{equation}
which shows the relation between $\varphi$ and $\alpha$ with certain tilting angle $\theta$.

\begin{figure}[t!]
	\includegraphics[width=0.65\linewidth]{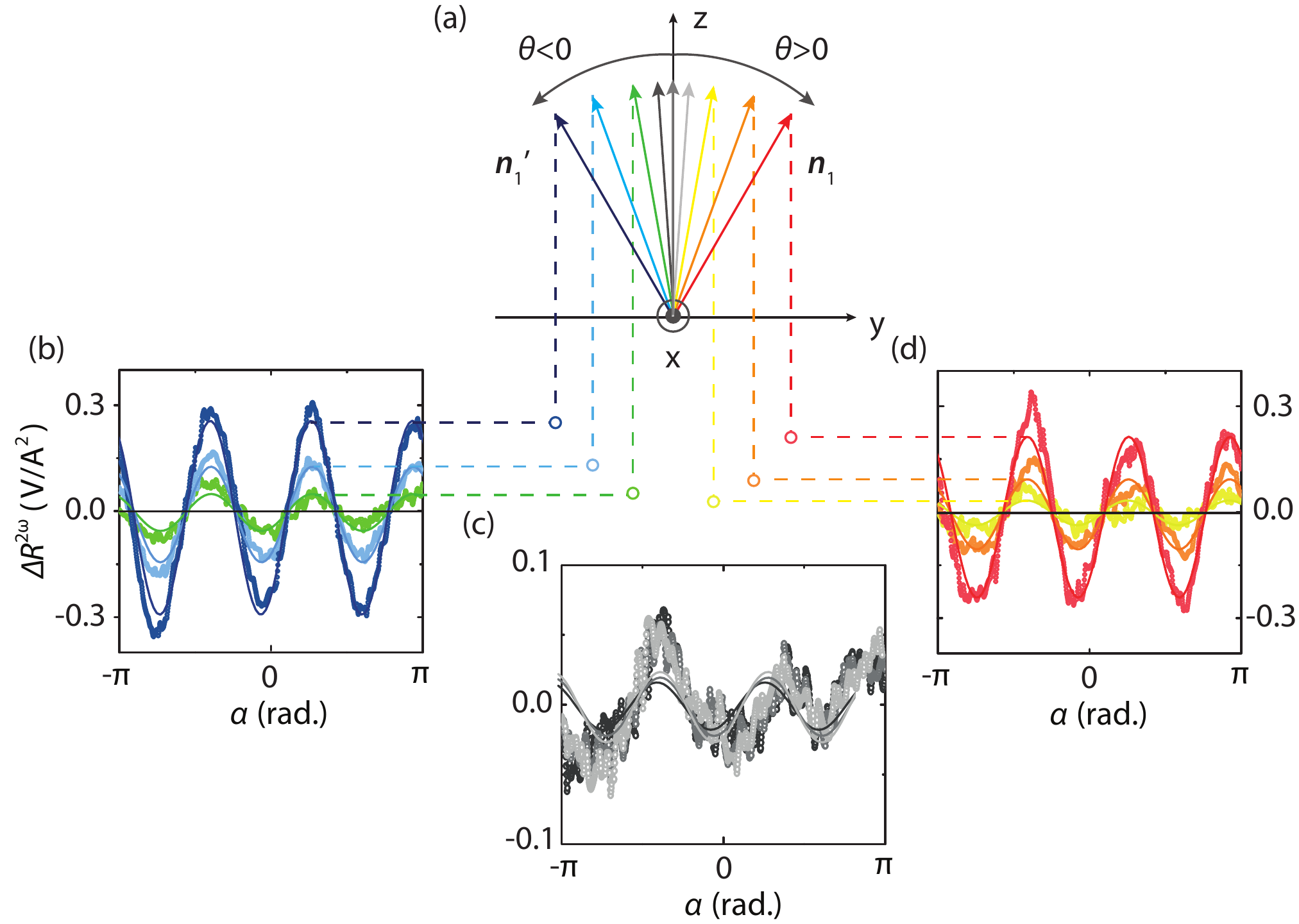}
	\caption {Tilting angle $\theta$ dependent $2\pi/3$-period sinusoidal modulation (series I, sample D, device 2 with $B=40\,$mT). (a) Normal vectors of the sample plane $\textbf{\textit{n}}_{1}$ and $\textbf{\textit{n}}_{1}^{\prime}$ with different tilting angles $\theta$ presented in different colors. The sign of $\theta$ is defined that for $\textbf{\textit{n}}_{1}$, $\theta>0$ and for $\textbf{\textit{n}}_{1}^{\prime}$, $\theta<0$. The $2\pi/3$-period sinusoidal components in the second harmonic signals with tilting angles of (b) $\theta=-30\,^{\textrm{o}}, -20\,^{\textrm{o}}, -10\,^{\textrm{o}}$, (c) $\theta=-4\,^{\textrm{o}}, 0\,^{\textrm{o}}, +4\,^{\textrm{o}}$ and (d) $\theta=10\,^{\textrm{o}}, 20\,^{\textrm{o}}, 30\,^{\textrm{o}}$. The colors of the $2\pi/3$-period sinusoidal modulation correspond to the color of the normal vectors in (a). We can obtain the magnitude of the $2\pi/3$-period sinusoidal modulation as a function of the tilting angle as shown by the colored circles here and in Fig. \ref{figS6_3}(a).}
	\label{figS6_2}
\end{figure}

\begin{figure}[h!]
	\includegraphics[width=0.9\linewidth]{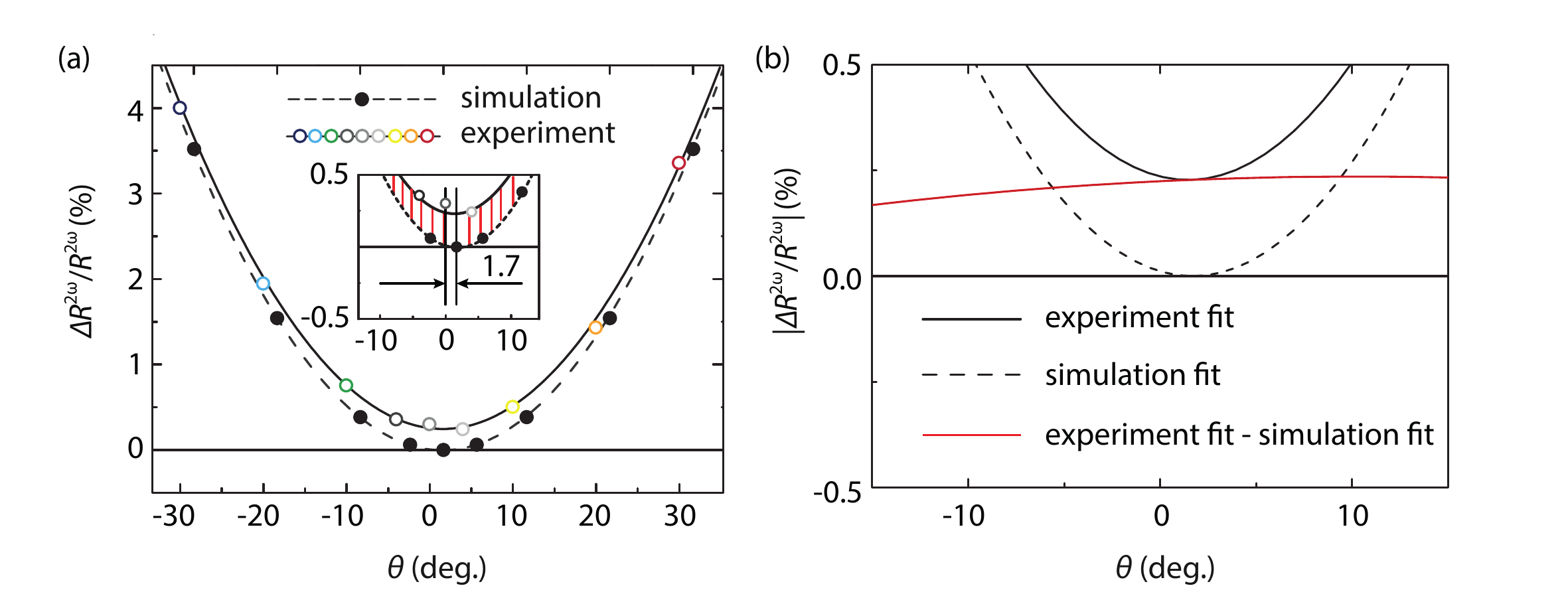}
	\caption {Comparison between the simulation and experiment for the sample tilting out-of-plane. (a) The ratio between the magnitude of the $2\pi/3$-period sinusoidal modulation $\Delta R^{2\omega}$ to the magnitude of the $2\pi$-period sinusoidal modulation $R^{2\omega}$ for the second harmonic signals as a function of tilting angle $\theta$. The simulation results are based on the model in Fig. \ref{fig_S6}, which only considers the influence from the out-of-plane tilting of the sample plane. A correction of 1.7$\,^\textrm{o}$  for angle $\theta$ is needed for the simulation results as shown in the inset, in order to fit the experimental data. The experimental results are from the data shown in Fig. \ref{figS6_2} with corresponding colors for different tilting angles. The error bars are within the size of the circles for all the data points. This is consistent within the accuracy of the alignment of $\pm 2\,^{\textrm{o}}$. The difference between the simulation and experiment in the small tilting angle regime is marked by the red striped area in the inset. The black dashed and solid lines are parabolic fits for the simulation and experimental data points, respectively. The zoomed-in figure of these fits are plotted in (b). The red solid line is the difference between the experiment and simulation fits, which corresponds to the amplitude of the MAMR around 0.2\%. The small MAMR amplitude measured by this device is due to the fact that the injector-to-detector spacing ($d=500\,$nm) is in the range where the sign and amplitude anomaly appears.}
	\label{figS6_3}
\end{figure}

Here, we assume that the in-plane magnetization aligns with $\textbf{\textit{B}}_{\textrm{in}}$. This is a reasonable assumption because the YIG films we used have a strong in-plane anisotropy, i.e. small in-plane saturation field and large out-of-plane saturation field ($\textbf{\textit{B}}_{\textrm{in}}^{\textrm{s}}<1.0\,\textrm{mT}$ and $\textbf{\textit{B}}_{\textrm{out}}^{\textrm{s}}>200.0\,\textrm{mT}$). We applied a magnetic field of 25 mT in the experiment described in this section. 

The magnon detection efficiency is modulated with $\varphi$ instead of $\alpha$. Therefore, $R^{2\omega}$ should have a $2\pi$-period sinusoidal oscillation as a function of $\varphi$. When the sample is tilted out-of-plane, i.e. $\theta\neq0$, the angle $\varphi$ is different from $\alpha$ with a relation depending on the magnitude and direction of the tilting angle $\theta$, for example as shown in Eq.$\ $(\ref{Eq_cosvarphi}). This gives rise to a $2\pi/3$-period sinusoidal modulation component in the second harmonic signals as a function of angle $\alpha$ as shown in Figs.$\,$\ref{fig_S6}(c) and (e). 

According to the simulation results shown in Fig.$\ $\ref{fig_S6}(e), the phase of this $2\pi/3$-period sinusoidal modulation depends on the tilting direction of the sample plane, i.e. the choice of the in-plane projection of the sample normal vector, for example, $\textbf{\textit{n}}_{1}$, $\textbf{\textit{n}}_{2}$ shown in Fig.$\ $\ref{fig_S6}(b), etc. This is also confirmed in the experimental results as shown in Figs.$\ $\ref{fig_S6}(d) and (f). We did the in-plane magnetic field angle dependent measurement while tilting the sample out-of-plane in a way that its normal vector has a certain in-plane projection, such as $\textbf{\textit{n}}_{1}$, $\textbf{\textit{n}}_{1}^{\prime}$ and $\textbf{\textit{n}}_{2}$. As shown in Fig.$\ $\ref{fig_S6}(f), comparing the cases of $\textbf{\textit{n}}_{1}$ and $\textbf{\textit{n}}_{2}$, the resulting $2\pi/3$-period sinusoidal modulations are out-of-phase to each other. By contrast, for the cases of $\textbf{\textit{n}}_{1}$ and $\textbf{\textit{n}}_{1}^{\prime}$, the $2\pi/3$-period sinusoidal modulations are in-phase. These features are consistent with the simulation results in Fig.$\ $\ref{fig_S6}(e).

The magnitude of the $2\pi/3$-period sinusoidal modulation depends on how much the sample tilts out-of-plane, i.e. the magnitude of $\theta$. In the experiment, we varied the angle $\theta$ by tuning the sample holder according to an angular scale with accuracy of $\pm2\,^{\textrm{o}}$. Figure \ref{figS6_2} shows different tilting angles and corresponding $2\pi/3$-period sinusoidal modulations from the measurement. The magnitude of the $2\pi/3$-period sinusoidal modulation from both simulation and experiment as a function of the tilting angle is summarized in Fig.$\ $\ref{figS6_3}. To fit the experimental data, especially in the large tilting angle regime, we add a correction of 1.7 degree for angle $\theta$ in the simulation results. The necessity of the small angle correction for the simulation also suggests that the flat sample tuned based on the scale is not really flat with an inaccuracy of less than 2 degree. 

Here, the simulation only takes into account the influence of sample tilting out-of-plane. However, in Fig. \ref{figS6_3}(a), there is a discrepancy between experiment and simulation. If the $2\pi/3$-period sinusoidal modulation were purely induced by the sample tilting out-of-plane, it would vanish when the sample is completely in-plane as shown in the simulation results. However, this is not the case here. As shown in Fig. \ref{figS6_3}(b), by varying $\theta$ from $-10\,^{\textrm{o}}$ to $10\,^{\textrm{o}}$, the variation of the difference between experiment and simulation for $\Delta R^{2\omega}/R^{2\omega}$ is less than 20\%. This difference between experiment and simulation corresponds to the magnitude of the MAMR feature for this device with $d=500\,$nm. It clearly proves that the MPHE and MAMR characteristic features, i.e. the $2\pi/3$-period sinusoidal modulation for the second harmonic signals, are not due to the sample out-of-plane tilting.

\subsection{\large VI. Sign and magnitude of the MPHE and MAMR}
We can determine if $\sigma^{\textrm{m}}_{\parallel}>\sigma^{\textrm{m}}_{\perp}$ or $\sigma^{\textrm{m}}_{\parallel}<\sigma^{\textrm{m}}_{\perp}$ for the magnon transport from the sign of the MPHE and MAMR signals, denoted as "$+$" and "$-$", respectively. In all the MPHE and part of the MAMR measurements, we observe "$-$". However, "$+$" is also obtained for some MAMR measurements with injector to detector spacing in a certain range. Here, we summarize the sign obtained from different devices and samples in Table \ref{tab:table2}, which corresponds to the data in Fig. $\textcolor{blue}{4}$.

\begin{figure}[t!]
	\includegraphics[width=0.35\linewidth]{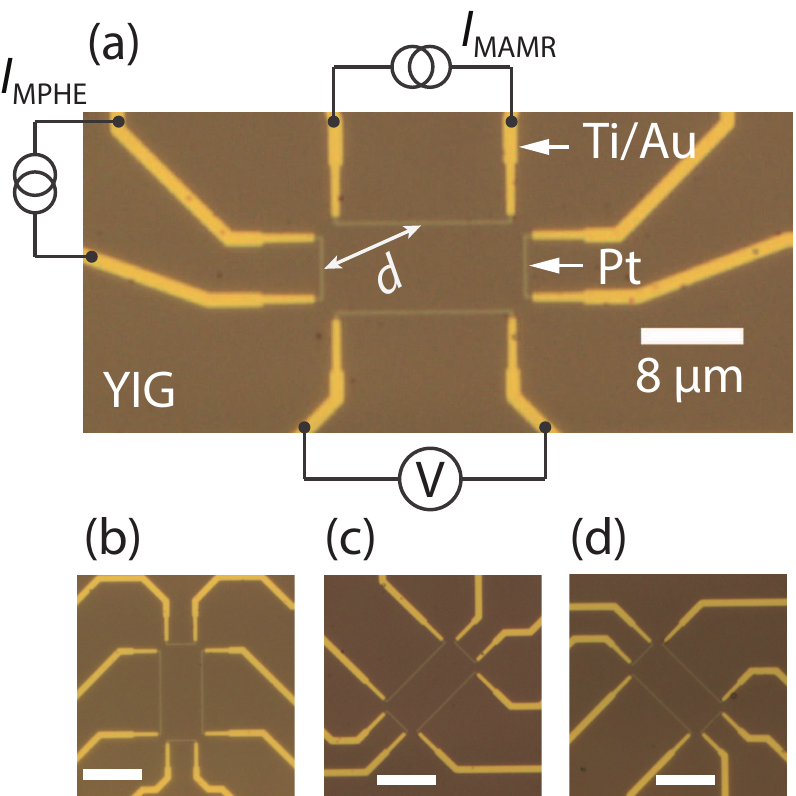}
	\caption {Optical images of the modified MPHE devices on sample E. (a) Both MPHE and MAMR measurement can be resolved with corresponding measurement configuration, i.e. MPHE ($I_{\textrm{MPHE}}$-$V$) and MAMR ($I_{\textrm{MAMR}}$-$V$). (b)-(d) The same device structure patterned on sample E with different orientation with respect to the crystalline orientation of the YIG (111) surface. Injector-to-detector distance $d$ for the MPHE device is defined as the spacing between the middle points of the injector and detector as shown in (a). The scale bars in (b)-(d) represent 10$\,\mu$m.}
	\label{MPHE_MAMR}
\end{figure}

\begin{table}[b!]
	\caption{\label{tab:table2}Summary of the sign measured with different geometries on different samples and devices on YIG thin film.}
	\begin{ruledtabular}
		\begin{tabular}{ccccccccc}
			Series (I or II)\footnotemark[1] 
			&Harmonic&I &I &I & I &II & II & II\\
			Samples (A-G)
			&n$\omega$&A &B &C & D &E & F & G\\
			Number of measured devices&(n$=$1 or 2)& 4 & 1 & 1 & 6 & 21 & 1 & 1\\
			\hline
			\multirow{2}{*}{Number of measured MPHE devices (sign\footnotemark[2])} &1$\omega$ &0&1($-$)&1($-$)&0&5($-$)&0&0\\
			&2$\omega$ & 4($-$) & 1($-$) & 1($-$) & 2($-$) & 13($-$)&0&0\\
			\multirow{2}{*}{Number of measured MAMR devices (sign)} &1$\omega$ &0&0&0&4($+$)&2($-$)3($+$)&1($-$)&0\\
			&2$\omega$ &0&0&0&4($+$)&5($-$)3($+$)&1($-$)&1($-$)

		\end{tabular}
	\end{ruledtabular}
	\footnotetext[1]{Series I are devices on 100 nm thick YIG, while series II are on 200 nm thick YIG.}
	\footnotetext[2]{Sign "$+$" indicates $\sigma^{\textrm{m}}_{\parallel}>\sigma^{\textrm{m}}_{\perp}$, while "$-$" represents $\sigma^{\textrm{m}}_{\parallel}<\sigma^{\textrm{m}}_{\perp}$.}
\end{table}

As can be seen in Table \ref{tab:table2}, the measured devices are patterned on two series of YIG samples with thickness of 100$\,$nm (series I) and 200$\,$nm (series II). 7 samples are labelled by letter from A to G. On each sample, multiple devices are patterned. In total, 35 devices are measured, including both MPHE and MAMR measurement devices. For each device, we can simultaneously measure the first ($1\omega$) and second ($2\omega$) harmonic signals, which always show the same sign. However, it is harder to resolve the characteristic $\pi/2$-period oscillation for the first harmonic signals, because the signal-to-noise ratio is generally smaller than that for the second harmonic signals. Therefore, we obtain only the second harmonic signals for some MAMR and MPHE measurements.  

All the MPHE measurements show "$-$" sign, i.e. $\sigma^{\textrm{m}}_{\parallel}<\sigma^{\textrm{m}}_{\perp}$. For MAMR measurement, we obtain the same sign for the devices on sample F and G where the injector to detector spacing $d$ is $200\,$nm whereas the opposite sign on sample D is observed ($d=500, 600, 800\,$nm and $1\,\mu$m). On sample E, MAMR devices show "$+$" for $d=600\,$nm but "$-$" for distance of $1\,\mu$m and $6\,\mu$m. In Fig. $\textcolor{blue}{4}$, we summarize the magnitude and sign of the MAMR as a function of the injector-to-detector distance. The anomaly of the sign for the devices with injector-to-detector spacing in certain range is not fully understood yet. We believe that this may be related to the sign reversal of $R^{2\omega}$ by increasing $d$ for YIG films with certain thickness and the thickness dependent behavior for the $R^{1\omega}$ \cite{PhysRevB.94.174437}.

Nevertheless, "$-$" is observed with the devices on sample E as shown in Fig. \ref{MPHE_MAMR} where we can measure MPHE and MAMR in the same region of YIG. Compared with the device structure for MPHE measurement in Fig. $\textcolor{blue}{1}$(a), the modified  MPHE structures in Fig. \ref{MPHE_MAMR} have two parallel strips relatively longer. With the longer stripes, we are able to resolve the small signals of $\Delta R^{1\omega}$ and $\Delta R^{2\omega}$ for the MAMR with the injector-to-detector spacing of $7\,\mu$m, since both first and second harmonic signals scale with the length of the strip. Besides, by patterning the devices on the same sample with the same structures but different orientations as shown in Figs. \ref{MPHE_MAMR}(b)-(d), we observe the same sign of "$-$". This indicates that the sign change is not caused by the device orientation with respect to the crystalline direction. However, to study the influence of the $C_{3}$ symmetry surface, more orientations should be systematically checked in the future.

\subsection{\large VII. Reciprocity and linearity of the MPHE and MAMR}

\begin{figure}[h!]
	\includegraphics[width=0.6\linewidth]{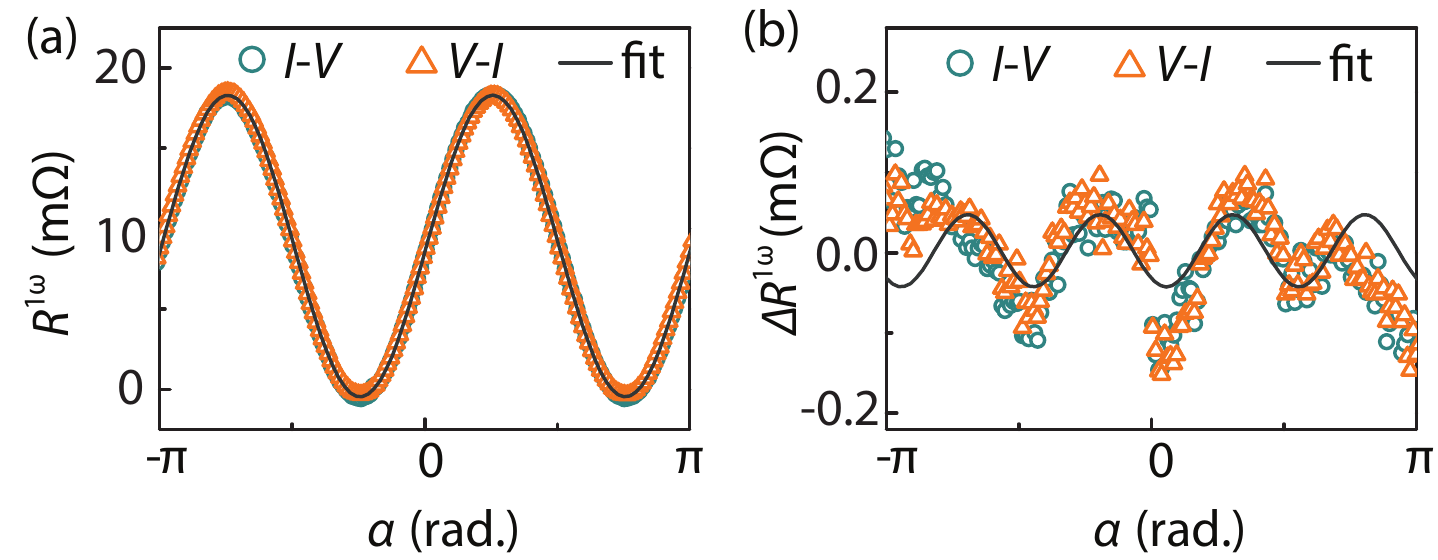}
	\caption {Reciprocity of $R^{1\omega}$  and $\Delta R^{1\omega}$ for the MAMR measurement with injector to detector spacing of 600$\,$nm (series II, sample E, device 19). (a) $R^{1\omega}$ and (b) $\Delta R^{1\omega}$ as a function of angle $\alpha$ for the $I$-$V$ and $V$-$I$ configurations. }
	\label{reciprocity}
\end{figure}

\begin{figure}[h!]
	\includegraphics[width=0.6\linewidth]{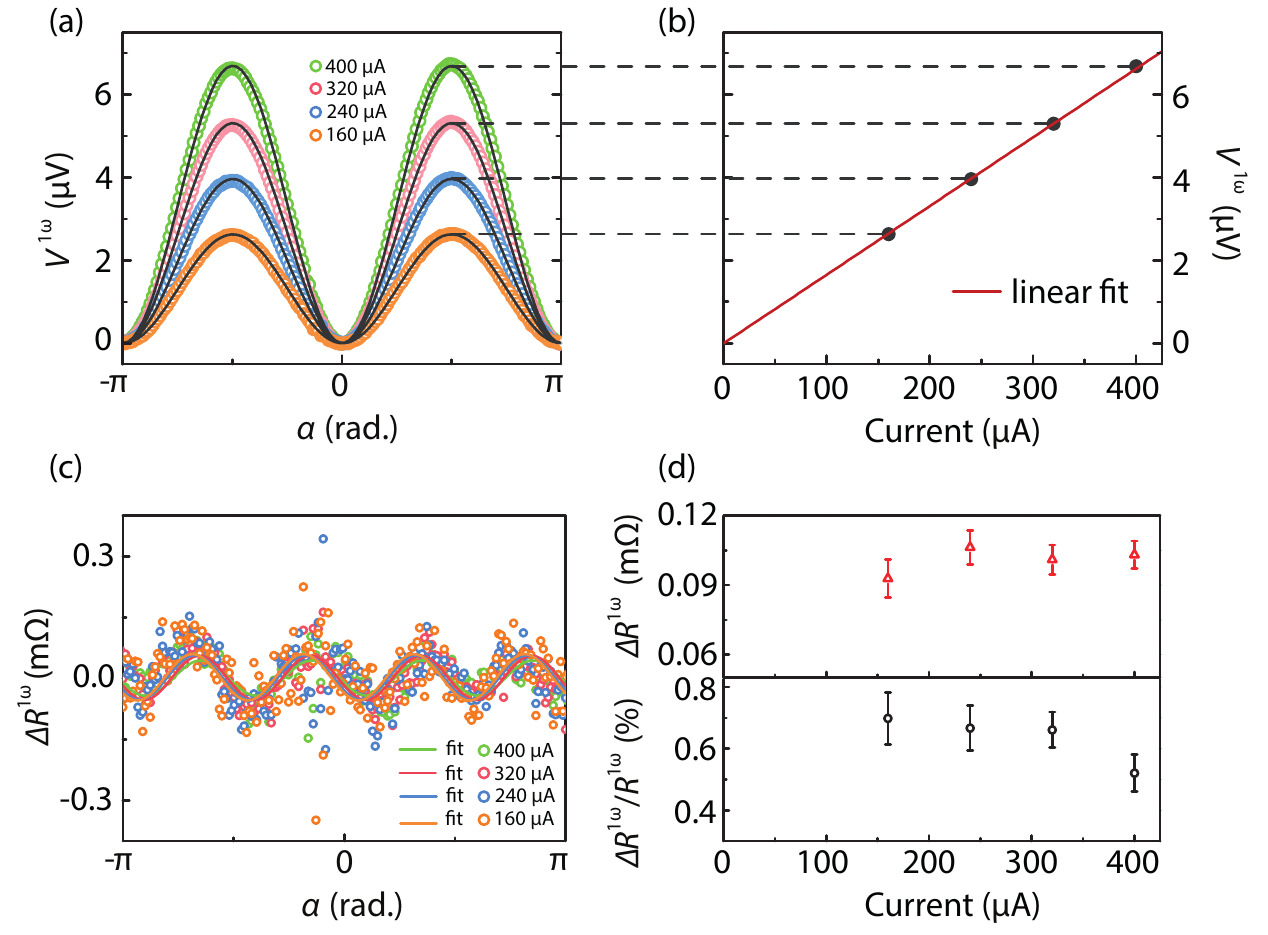}
	\caption {Linearity of the first harmonic signals for the MAMR measurement with injector to detector spacing of 600$\,$nm (series II, sample E, device 18). (a) Angular dependences of $V^{1\omega}$ with different excitation currents. Black solid lines are $\pi$-period sinusoidal fits. (b) Peak-to-peak amplitudes of $V^{1\omega}$ as a function of the current. The error bar is smaller than the radius of the data points. The red solid line is a linear fit through the data, showing the linearity of $V^{1\omega}$. (c) Angular dependence of $\Delta R^{1\omega}$, i.e. the residues of the $\pi$-period sinusoidal fits for $R^{1\omega}=V^{1\omega}/I$, for different excitation currents. The corresponding colored solid lines are the $\pi/2$-period sinusoidal fits. (d) Magnitude of $\Delta R^{1\omega}$ and $\Delta R^{1\omega}/R^{1\omega}$ as a function the current. They do not depend on the current within the experimental uncertainty, which proves the linearity of the MAMR.}
	\label{linearity}
\end{figure}

We perform the reciprocity measurement by reversing the role of injector and detector as shown in Fig. \ref{reciprocity} on a MAMR device with $d=600\,$nm (series II, sample E, device 19). This is to verify that the magnon injection and detection are in the linear regime where Onsager reciprocity holds. We find that $R^{1\omega}_{I-V}=18.8 \pm0.1\,$m$\Omega$ and $R^{1\omega}_{V-I}=18.7 \pm0.1\,$m$\Omega$, while $\Delta R^{1\omega}_{I-V}=0.08 \pm0.01\,$m$\Omega$ and $\Delta R^{1\omega}_{V-I}=0.09 \pm0.01\,$m$\Omega$. Since $R^{1\omega}_{I-V}(B)=R^{1\omega}_{V-I}(-B)$ and $\Delta R^{1\omega}_{I-V}(B)=\Delta R^{1\omega}_{V-I}(-B)$, we confirm the Onsager reciprocity holds for both $R^{1\omega}$ and $\Delta R^{1\omega}$ within the experimental uncertainty \cite{cornelissen2015long}. Moreover, linearity has been confirmed by measuring the first harmonic signals with various currents on a MAMR device with $d=600\,$nm as shown in Fig. \ref{linearity}.

\bibliography{reference}